\documentclass[epj]{svjour}
\usepackage{graphics}
\usepackage{amssymb,amsmath}
\usepackage{graphicx}
\usepackage[usenames]{color}
\usepackage{bbold}
\usepackage{slashed} 

\newcommand{\blue}[1]{\textcolor{blue}{#1}}

\newcommand{\bea}{\begin{eqnarray}}
\newcommand{\eea}{\end{eqnarray}}
\newcommand{\ba}{\begin{align}}
\newcommand{\ea}{\end{align}}
\newcommand{\bma}{\begin{pmatrix}}
\newcommand{\ema}{\end{pmatrix}}

\usepackage[percent]{overpic}
\usepackage{overpic}

\begin{document}
\title{Neutrinos propagating in curved spacetimes}
\author{Fan Zhang\inst{1}\inst{2} 
}                     
%
%
\institute{Gravitational Wave and Cosmology Laboratory, Department of Astronomy, Beijing Normal University, Beijing 100875, China \and Department of Physics and Astronomy, West Virginia University, PO Box 6315, Morgantown, WV 26506, USA}
\date{Received: date / Revised version: date}
%
\abstract{
In the Dirac-Weyl equation that describes massless neutrino propagation in the minimal Standard Model, the ($2\times 2$ equivalence of the) gamma matrices convert Weyl spinors into spacetime tensors, and vice versa. They can thus be regarded as amalgamations of three different types of mappings, one that connects particle spinors directly forming representations to internal gauge symmetries to their spacetime counterparts that are embodied by null flags, another that translates the spacetime spinors into their corresponding tensors expressed in an orthonormal tetrad, and finally a purely tensorial transformation into the coordinate tetrad. The splitting of spinors into particle and spacetime varieties is not usually practised, but we advocate its adoption for better physical clarity, in terms of distinguishing internal and spacetime transformations, and also for understanding the scattering of neutrinos by spacetime curvature. We construct the basic infrastructure required for this task, and provide a worked example for the Schwarzschild spacetime. Our investigation also uncovers a possible under-determinacy in the flavoured Dirac-Weyl equation, which could serve as a new incision point for introducing flavour oscillation mechanisms. 
} 
\maketitle
\section{Introduction} 
Neutrino phenomenology is frequently coupled with strong gravitational fields. For example, neutrinos are expected to play a vital role in reigniting the stalled explosion during core collapse supernovae \cite{Colgate:1966ax,1966CaJPh..44.2553A}. Therefore, it is important to understand the nuts and bolts of how the spacetime curvature scatters neutrinos. More generally, combining particle physics with classical gravity has yielded important insights such as the Hawking radiation \cite{Hawking:1974sw}, so a physically complete procedure for transcribing flat spacetime expressions into curved regions is of outstanding utility. 

\begin{figure}[b!]
\setlength{\unitlength}{1cm}
\begin{picture}(8.5,3.5)(0,0)
\linethickness{1.pt}
\put(1,3.35){\makebox(0,0){(coord)}}
\put(1,3.7){\makebox(0,0){Spacetime}}
\put(1,2.8){\makebox(0,0){\blue{[simple idx]}}}
\put(4,3.35){\makebox(0,0){(ortho)}}
\put(4,3.7){\makebox(0,0){Spacetime}}
\put(4,2.8){\makebox(0,0){\blue{[tilde idx]}}}
\put(7,3.35){\makebox(0,0){(internal)}}
\put(7,3.7){\makebox(0,0){Particle}}
\put(7,2.8){\makebox(0,0){\blue{[hat idx]}}}
\put(1,1.5){\makebox(0,0){Tensor}}
\put(4,1.5){\makebox(0,0){Tensor}}
\put(4,0.2){\makebox(0,0){Spinor}}
\put(7,0.2){\makebox(0,0){Spinor}}
\linethickness{1pt}
\put(1.7,1.5){\line(3,0){1.55}}
\put(4.7,0.2){\line(3,0){1.55}}
\put(1.3,1.2){\line(5,-2){2}}
\put(4,0.45){\line(0,1){0.8}}
\put(1,1.8){\line(0,1){0.4}}
\put(1,2.18){\line(1,0){6}}
\put(7,0.5){\line(0,1){1.7}}
\put(6.8,1.3){\makebox(0,0){$\gamma$}}
\put(5.5,0.4){\makebox(0,0){$S$ [1]}}
\put(2.5,1.7){\makebox(0,0){$a/b$ [3]}}
\put(4.5,0.9){\makebox(0,0){$\gamma_0$ [2]}}
\put(1.7,0.7){\makebox(0,0){$\tilde{\gamma}_0$}}
\end{picture}
\caption{A summary of relationships between quantities existing in the internal particle space and on the external spacetime, with the latter group splitting into those expressed under the coordinate tetrad and an orthonormal tetrad. The numbers over connecting lines correspond to the enumeration in the main text, and the blue texts in square brackets indicate how these different types of quantities are distinguished by their indices. }
\label{fig:FrameIllustration}
\end{figure}
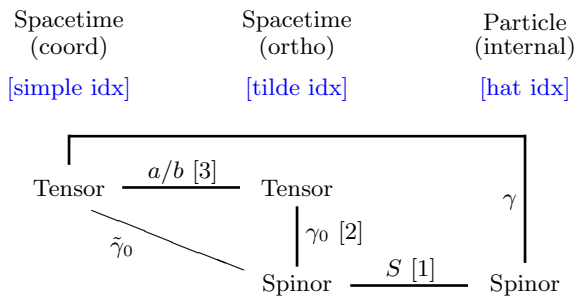

We begin our examination of neutrino propagation with a closer look at the $2\times 2$ gamma matrices (we misappropriate this nomenclature to describe the relevant Clifford generators, as alternative names such as Pauli matrices are more commonly associated with specific constant numerical values) that appear in e.g., the neutrino current density $j_{\alpha} = {\bar{\psi}}_{\hat{A'}}\gamma_{\alpha\, \hat{A}}{}^{\hat{A'}}{\psi}^{\hat{A}}$ (Greek letters denote spacetime indices, while capital Latin letters are used for spinor indices). 
This expression shows that the gamma matrices translate the directional information contained in the Weyl spinors into a regular spacetime vector, a task that can be accomplished in stages (see Fig.~\ref{fig:FrameIllustration} for notation): (1) one can define (see Sec.~\ref{sec:Rationale} below for details) particle spinors ${\psi}^{\hat{A}}$ that by themselves already form representations to the internal Standard Model (SM) gauge symmetries but are invariant under spacetime transformations, which are then mapped into spacetime spinors (changes under spacetime transformations) that are geometrically null flags or half null planes attached to null vectors \cite{Penrose1992}, via $\psi^{\tilde{B}}=S^{\tilde{B}}{}_{\hat{A}}{\psi}^{\hat{A}}$ ($S$ will be referred to as the decipherer); (2) The spacetime spinors are then mapped to tensorial quantities in an orthonormal tetrad using a set of constant soldering forms $\gamma_{0\, \tilde{A}\tilde{A}}{}^{\tilde{\beta}}$ ($*^1$, we use enumerated asterisks to mark statements for which further details are provided in the Appendix); (3) The tensors are further transplanted onto the coordinate tetrad via the vierbein 
$a^{\alpha}{}_{\tilde{\beta}}$ and its inverse $b_{\beta}{}^{\tilde{\alpha}}$. We thus have $\gamma_{\alpha \, \hat{A}}{}^{\hat{A'}}= \bar{S}^{-1}_{\tilde{B'}}{}^{\hat{A'}}\tilde{\gamma}_{0\, \alpha \, \tilde{B}}{}^{\tilde{B'}}S{}^{\tilde{B}}{}_{\hat{A}}$, 
where $\tilde{\gamma}_{0\, \tilde{A}\tilde{A'}}{}^{\alpha} \equiv a^{\alpha}{}_{\tilde{\beta}}\gamma_{0\, \tilde{A}\tilde{A'}}{}^{\tilde{\beta}}$. 

The first transformation is not usually incorporated into investigations of neutrino propagation, because the decipherer would be a simple constant in flat spacetimes that can be set to unity. Even in studies considering curved spacetimes (where the decipherer becomes variable), the gamma matrices usually only refer to $\tilde{\gamma}_{0}$ and the particle spinors are not invoked. In this paper, we introduce this extra layer of transformation (1), and show that despite being mathematically equivalent to the traditional treatments in the single flavour case, the additional step of peeling off a separate identity for the particle spinors, and utilizing the decipherer-enhanced gamma matrices $\gamma_{\alpha}$ (it is well known that the gamma matrices appearing in the Dirac-Weyl equation are not unique, so our additions do not conflict with any existing infrastructure), offers better conceptual clarity, as well as practical computational benefits (it removes the arbitrariness associated with the orthonormal tetrad choice, leading directly to physically meaningful quantities) and new avenues for modelling flavour oscillations (note this serendipitous find does not require curved spacetimes, although the underlying clues leading up to it are best revealed when gravity is present).  

\section{The benefits of a separate particle spinor} \label{sec:Rationale}
Algorithmically, although $\psi^{\hat{A}}$ and $\psi^{\tilde{A}}$ can be seen as the components of the same abstract spinor $\boldsymbol{\psi}$ decomposed onto two alternative dyad bases, so the decipherer can be seen simply as a dyad transformation, the two dyads are very different physical animals \footnote{In other words, the particle spinors we define below are not new entities in a mathematical sense, and all the familiar mathematical definitions and theorems regarding spinors still apply. If one does not wish to change anything with the existing theories, these particle spinors can be implemented simply via a spinor dyad transformation, leading to no alterations to the observables. They are however different in their physical significance, and thus can become useful when we attempt to expose the physical intuition beneath the mathematics. Aside from the more immediate benefits discussed in this paper, our ultimate goal for introducing them is to have them serve as conduits linking up with spacetime-coordinate-invariant measurements on particles in future works.}. The $\psi^{\tilde{0}}$ and $\psi^{\tilde{1}}$ are the components of $\boldsymbol{\psi}$ on the spacetime dyad basis equivalent to the arbitrary orthonormal tetrad (see $*^1$ for the precise mapping) that we choose to carry out spacetime computations in (the dyad bases do not satisfy any equation of motion). As it is simply the spinor version of the tetrad, it changes under spacetime transformations (e.g., the dyad bases are acted on by the spinor version of the Lorentz transformations $\text{SL}(2,\mathbb{C})$). Individual 
spinors decomposed onto this dyad, like vectors decomposed onto the tetrad, are thus regular spacetime entities (essentially null spacetime vectors, and more precisely null flags), to be acted on by spacetime transformations\footnote{Note though, a subset of spacetime spinor fields that satisfy particle propagation equations can serve as vectors in an abstract representation space. Internal gauge transformations can then act in this space, operating on the spacetime spinors as if they are abstract black box objects. The internal transformations never act directly on the two components of the spacetime spinor.} (as active counterparts to the same transformations that act on the spacetime dyad). 

On the other hand, we define $\psi^{\hat{0}}$ and $\psi^{\hat{1}}$ to be components of $\boldsymbol{\psi}$ on a physical dyad, whose bases are actual spinor \emph{fields} satisfying the evolution equations for propagating massless particles (the decipherer equation of motion derived later). These propagating fields are physical covariant quantities, thus not really altered by spacetime transformations, and should be regarded as avatars of whatever physical entities (more precisely the basis frame bases for these entities) that particle fields ultimately represent. Therefore, internal particle transformations can legitimately act on them. Alternatively stated, these propagating fields form an abstract solution space, which is rightfully the representation space of gauge symmetries (since they are by definition transformations of one solution into another). The expression $\psi^{\hat{A}}$ can then be elevated to essentially a vector in this abstract solution space, and the same internal transformations can act either actively on it (directly on the components of the particle spinor, without needing to construct additional representation space), or passively on the physical dyad. 

Previous literature (see e.g.~\cite{Brill:1957fx,Weinberg72,Birrell:1982ix}) effectively sets $S{}^{\tilde{B}}{}_{\hat{A}}$ to identity, and only allow $a^{\alpha}{}_{\tilde{\beta}}$ to be variable in a gravitational field. This does not lead to wrong predictions for observable quantities, as we have equalities such as ${\bar{\psi}}_{\tilde{B'}}\tilde{\gamma}_{0\, \alpha\, \tilde{B}}{}^{\tilde{B'}}{\psi}^{\tilde{B}}={\bar{\psi}}_{\hat{A'}}\gamma_{\alpha\, \hat{A}}{}^{\hat{A'}}{\psi}^{\hat{A}}$, so we can see those works as evolving 
${\psi}^{\tilde{B}}$ instead of the more rudimentary ${\psi}^{\hat{A}}$. However, without invoking ${\psi}^{\hat{A}}$, the physical picture can become muddled, and one may in rare occasions mistakenly conclude that internal gauge symmetries can be regarded as spacetime transformations. No such confusion would likely arise with the majority of internal gauge transformations that blend multiple different spinor fields, as new indices are introduced (to label these different fields, giving e.g., $\psi^{\tilde{B}\, i}$ where $1\leq i\leq n$ for some $n$) for the symmetry transformations to act on (e.g., as $\psi^{\tilde{B}\, i} \rightarrow \zeta{}^{j}{}_{i} \psi^{\tilde{B}\, i}$ for some gauge transformation $\zeta$), which makes clear the existence of an abstract representation space\footnote{This works in formally the same way with particle spinors, via $\psi^{\hat{A}\, i} \rightarrow \zeta{}^{j}{}_{i} \psi^{\hat{A}\, i}$, i.e., although $\zeta$ can also act on each particle spinor field separately, they only form trivial representations individually. Nevertheless, unlike with spacetime spinors, the overall representation is now truly $2n$ dimensional.}. Instances do occur however, when the representation is one dimensional, so the gauge transformations appear to operate directly on a single spinor field, without involving any new indices to signal the existence of the abstract representation space. Take the simplest case of the electromagnetic gauge transformation that introduces an overall phase factor onto a single spinor (with a numerical value of one for charge-neutral particles, but a transformation is nonetheless applied, just with the identity element of the symmetry group; this issue of separating spinor roles is also not limited to neutrinos, so nonvanishing charges are also relevant for the present discussion). Without ${\psi}^{\hat{A}}$, this transformation would operate directly on ${\psi}^{\tilde{B}}$ as ${\psi}^{\tilde{B}} \rightarrow e^{i\phi} {\psi}^{\tilde{B}}$ for some $\phi \in \mathbb{R}$, and can easily be misconstrued as effecting a rotation of the null flag that embodies ${\psi}^{\tilde{B}}$ (changing the direction of the half null plane attached to the null vector), even though it is clearly not intended to be a spacetime transformation. Such confusions can be avoided if the said transformation is applied to ${\psi}^{\hat{A}}$ instead. Since then it would be clear that it is really an abstract operator whose actions are explicated by transformations of a different type of spinor components serving a different role, as decomposition coefficients of a vector in the aforementioned abstract solution space, thereby shifting it into a different but physically equivalent vector/solution (the definition of an internal gauge transformation). 

It is worthwhile emphasizing again that the gauge transformations can act directly on the components of the single particle spinor field. In contrast, as already alluded to, to have internal transformations acting on a single spacetime spinor, one would have to build an extra layer of infrastructure where an abstract one-dimensional representation space is first constructed with spacetime spinor fields being vectors in it (it is important that the vectors in this representation space are solutions to particle propagation equations, just like the dyad bases for the physical spinors, as the gauge transformations are symmetries to these equations), and then the transformations can act on this representation space instead. For example, the $e^{i\phi}$ above really acts on the coefficient of $\psi^{\tilde{B}}$ seen as an abstract vector in this representation space, and not on its two components $\psi^{\tilde{0}}$ and $\psi^{\tilde{1}}$ (one would perhaps more accurately write the transformation as $1\times \psi^{\tilde{B}} \rightarrow (e^{i\phi}\times 1)\times \psi^{\tilde{B}}$). Failure to realize this rather subtle point would lead to the aforementioned confusions. On the other hand, the phase transformation does act directly on $\psi^{\hat{0}}$ and $\psi^{\hat{1}}$, so there will be no chance for confusion with the expression $\psi^{\hat{A}}\rightarrow e^{i\phi}\psi^{\hat{A}}$. This desirable improvement ultimately comes from the fact that particle spinors by themselves already form representations to the internal symmetries, and are thus immediate handles on the physical particles that the internal transformations directly crank. In other words, although defining the particle spinors may appear as a somewhat redundantly pedagogical exercise, it should prove useful when we further excavate the physical interpretations of the geometrical languages utilized by the SM. One example of potential new physics that can be obtained this way arises when incorporating flavours, which we discuss in Sec.~\ref{sec:Flavour}. Another is that internal transformations of the form $\psi^{\hat{A}}\rightarrow \zeta^{\hat{B}}{}_{\hat{A}}\psi^{\hat{A}}$ (i.e. acting on a single spinor yet differently on its two components) are now possible should the need for them arises.  

In addition, adopting ${\psi}^{\hat{A}}$ and the associated dynamic decipherer field also allows for sanitizing the study of scattering by spacetime curvature, removing the spurious effects resulting from the arbitrariness of the spacetime dyad (equivalently, the tetrad). Namely, this dyad has an arbitrary variation as we move about spacetime, which has nothing to do with gravity but nevertheless manifests in the same way in the spin connection, blending with and obstructing our extraction of the real curvature scattering effects. In contrast, the physical dyad, consisting of actual propagating fields, do not suffer from this complication. 
Some basic infrastructure are needed if we are to reap the benefits of the multiple spinor personalities. In particular, we need to have at hand the propagation equations of the physical dyad bases (they are simply $S{}^{\tilde{B}}{}_{\hat{0}}$ and $S{}^{\tilde{B}}{}_{\hat{1}}$). We turn to their derivation next. 

\section{Decipherer equation of motion} 
The defining relation 
$\bar{\gamma}_{\alpha \, \hat{A'}}{}^{\hat{C}}\gamma_{\beta\,\hat{B}}{}^{\hat{A'}}+\bar{\gamma}_{\beta\,\hat{A'}}{}^{\hat{C}}\gamma_{\alpha\,\hat{B}}{}^{\hat{A'}} = -g_{\alpha \beta} \epsilon_{\hat{B}}{}^{\hat{C}}$ ($g$ is the spacetime metric and $\epsilon$ is the spinor ``metric" that raises or lowers indices via $\kappa_{\hat{B}}=\kappa^{\hat{A}}\epsilon_{\hat{A}\hat{B}}$ and $\kappa^{\hat{A}}=\epsilon^{\hat{A}\hat{B}}\kappa_{\hat{B}}$) 
for the gamma matrices to generate a Clifford algebra is satisfied by any invertible decipherer field ($*^2$). To determine its actual value though, we need to derive the decipherer equation of state (DEOM)($*^3$). To this end, we note that the Dirac-Weyl equation $\gamma_{\hat{A}\hat{A'}}{}^{\alpha}\left(\epsilon_{\hat{B}}{}^{\hat{A}}\partial_{\alpha} + \Gamma_{\alpha\, \hat{B}}{}^{\hat{A}} \right) {\psi}^{\hat{B}} = 0$ can be written in the purely spinorial form of $\nabla_{\hat{A}\hat{A'}}\psi^{\hat{A}}=0$, which is concise and appealing but somewhat unnatural, since the derivatives carry particle instead of spacetime indices. It is reasonable then to assert that it is derived from a more fundamental equation $\nabla^{\tilde{A}\tilde{A'}}(S^{-1}_{\tilde{A}}{}^{\hat{B}}\psi_{\hat{B}}) = 0$, which is the standard freely propagating massless field equation that also governs the spinor versions of the Weyl curvature tensor and the Faraday tensor (Ref.~\cite{Penrose1992} Eqs.~4.10.9 and 5.1.57). Applying the product rule ($*^4$), we then obtain the desired DEOM $\nabla^{\tilde{A}\tilde{A'}}S^{-1}_{\tilde{A}}{}^{ \hat{E}} = 0$, which is also that of a massless freely propagating field (see last section). 

To facilitate computations, we make the decomposition 
$S^{-1}{}_{\tilde{D}}{}^{\hat{E}} = C^{\hat{E}}_{\o}{}\o_{\tilde{D}}+C^{\hat{E}}_{\iota}\iota_{\tilde{D}}$ onto the spacetime spinor dyad $(\o_{\tilde{D}},\iota_{\tilde{D}})$ corresponding to an orthonormal tetrad. The DEOM then takes up the more explicit form of ($*^5$)
\begin{align} \label{eq:EOMCompT}
(\epsilon-\rho)C^{\hat{E}}_{\o}+(\pi-\alpha)C^{\hat{E}}_{\iota}+DC^{\hat{E}}_{\o}+\delta'C^{\hat{E}}_{\iota}&=0\,,\notag \\
(\beta-\tau)C^{\hat{E}}_{\o}+(\mu-\gamma)C^{\hat{E}}_{\iota}+\delta C^{\hat{E}}_{\o}+D'C^{\hat{E}}_{\iota}&=0\,,
\end{align}
where the Greek letters are the Newman-Penrose (NP) \cite{1962JMP.....3..566N} spin coefficients. Note that only spacetime spinors require covariant derivatives, so e.g., coefficients $C_{\o}^{\hat{E}}$ should be treated as two scalar fields, and the NP intrinsic derivatives $D$, $D'$, $\delta$, $\delta'$ in Eq.~\eqref{eq:EOMCompT} reduce to directional partial derivatives. 

\section{Adding flavours} \label{sec:Flavour}
Because neutrinos can switch flavours during propagation, it is useful to introduce generations into their propagation equations and examine the implications for potential oscillation mechanisms. 
The exercise of introducing flavours is generally carried out via ad hoc procedures, given the mysteries surrounding generational physics (epitomized by the famous ``who ordered that?" quip by Rabi). Our distinguishing between the particle and spacetime spinors provides additional flexibilities as to how this can be done, beyond simply making three copies of everything.  

In the SM, the flavours appear through an additional flavour index $\hat{f}_0$, so the particle spinors become $\psi^{\hat{A}\, \hat{f}_0}$. Furthermore, generational physics is only relevant to particle spinors, while the spacetime spinors are always those null flags or more colloquially ``square-roots of null vectors'' \cite{Wald} that there is only one copy of. So flavour indices can only appear on the particle side of the decipherer, which becomes $S{}^{\tilde{B}}{}_{\hat{A}\, \hat{f}_0}$. The inverse $S^{-1}$ is now defined by the simultaneous satisfaction of both the right- and left-inverse conditions ${S}{}^{\tilde{B}}{}_{\hat{C}\, \hat{f}_0}{S}^{-1}_{\tilde{A}}{}^{\hat{C} \, \hat{f}_0}
= \epsilon_{\tilde{A}}{}^{\tilde{B}}$ and ${S}^{-1}_{\tilde{B}}{}^{\hat{A} \, \hat{f}_1}{S}{}^{\tilde{B}}{}_{\hat{C}\, \hat{f}_0}= \epsilon_{\hat{C}}{}^{\hat{A}}\mathbb{F}_{\hat{f}_0}{}^{\hat{f}_1}$, where $\mathbb{F}$ is the flavour space metric. Such an $S^{-1}$ only exists ($*^6$) if the flavour content of the decipherer factorizes out in such a way that we can write 
$S{}^{\tilde{B}}{}_{\hat{A}\, \hat{f}_0}=\mathcal{V}_{\hat{f}_0}S_{e}{}^{\tilde{B}}{}_{\hat{A}}$ where $\mathcal{V}=(1,\mathcal{F}_{\mu},\mathcal{F}_{\tau})/\sqrt{3}$. The $\mathcal{F}$s can in principle be scalar fields, but are restricted by the flavoured DEOM $\nabla^{\tilde{A}\tilde{A'}}S^{-1}_{\tilde{A}}{}^{\hat{E} \, \hat{f}_0} =0$ ($*^7$) to constants. Their values ought to be decided by experiments, but by demanding that the decipherer maximally preserves orthonormality while mapping between particle and spacetime spinor dyads in a flat spacetime, we can narrow down their candidate pool into ($*^8$) $\mathcal{F}_{\mu} = e^{i\phi_\mu}\,, \, \mathcal{F}_{\tau} = e^{i\phi_\tau}$.

The right-inverse condition gives $S^{-1}_{\tilde{A}}{}^{\hat{B}\, \hat{f}_0}=S^{-1}_{e\, \tilde{A}}{}^{\hat{B}}\mathcal{V}^{\hat{f}_0}$, where $\mathcal{V}^{\hat{f}_0} = (1,\mathcal{F}^{-1}_{\mu},\mathcal{F}^{-1}_{\tau})^T/\sqrt{3}$ (under the $\mathcal{F}$ fixing above, raising the index on $\mathcal{V}$ with $\mathbb{F}$ is equivalent to the more familiar conjugated transpose), while the left-inverse condition implies that $\mathbb{F}$ must be a projection operator, specifically $\mathbb{F}_{\hat{f}_0}{}^{\hat{f}_1}=\mathcal{V}_{\hat{f}_0}\mathcal{V}^{\hat{f}_1}$. The gamma matrices that carry information about the spacetime metric (through $a^{\alpha}{}_{\tilde{\beta}}$) become 
\begin{align} \label{eq:GDecompose}
\gamma_{\alpha \, \hat{A}\, \hat{f}_0}{}^{\hat{A'}\, \hat{f}_1} 
&= b_{\alpha}{}^{\tilde{\beta}}\bar{S}^{-1}{}_{\tilde{B'}}{}^{\hat{A'} \, \hat{f}_1}{}{\gamma}_{0\, \tilde\beta \, \tilde{B}}{}^{\tilde{B'}}S{}^{\tilde{B}}{}_{\hat{A} \, \hat{f}_0}\notag \\
&= \left[b_{\alpha}{}^{\tilde{\beta}}\bar{S}^{-1}_{e\, \tilde{B'}}{}^{\hat{A'}}{\gamma}_{0\, \tilde\beta \, \tilde{B}}{}^{\tilde{B'}}S_{e}{}^{\tilde{B}}{}_{\hat{A}}\right]\left(\mathcal{V}_{\hat{f}_0}\bar{\mathcal{V}}^{\hat{f}_1}\right)\notag \\
&\equiv \left[\gamma_{e\, \alpha \, \hat{A}}{}^{\hat{A'}}\right]\left(\frac{\mathbb{G}_{\hat{f}_0}{}^{\hat{f}_1}}{3}\right)\,,
\end{align}
and determine a spin connection $\Gamma_{\alpha \, \hat{B}}{}^{\hat{A}}$ for parallelly transporting particle spinors through the metricity condition 
\begin{align} \label{eq:MetricityFullT}
0=\frac{\partial\gamma_{\alpha\, \hat{B}\, \hat{f}_0}{}^{\hat{B'}\, \hat{f}_1}}{\partial x^{\beta}}
- \Gamma^{\xi}_{\alpha \beta} \gamma_{\xi\, \hat{B}\, \hat{f}_0}{}^{\hat{B'}\, \hat{f}_1}
+ \bar{\Gamma}_{\beta\, \hat{D'}\, \hat{f}_2}{}^{\hat{B'}\, \hat{f}_1}\gamma_{\alpha\, \hat{B}\, \hat{f}_0}{}^{\hat{D'}\, \hat{f}_2}
-\gamma_{\alpha\, \hat{C}\, \hat{f}_3}{}^{\hat{B'}\,\hat{f}_1}\Gamma_{\beta\, \hat{B}\, \hat{f}_0}{}^{\hat{C}\, \hat{f}_3}
\,.
\end{align}
When interacting with $\mathbb{G}$, $\mathbb{F}$ exhibits behaviours typical of a metric
\bea \label{eq:GIdentities}
\bar{\mathbb{F}}_{\hat{f}_3}{}^{\hat{f}_1}\mathbb{G}_{\hat{f}_0}{}^{\hat{f}_3}=\mathbb{G}_{\hat{f}_0}{}^{\hat{f}_1}=\mathbb{G}_{\hat{f}_2}{}^{\hat{f}_1}\mathbb{F}_{\hat{f}_0}{}^{\hat{f}_2}\,,
\eea
thanks to which Eq.~\eqref{eq:MetricityFullT} admits closed form solutions ($*^9$)
\begin{align} \label{eq:SpinConnectionFactor}
\Gamma_{\beta \, \hat{B}\, \hat{f}_0}{}^{\hat{C}\, \hat{f}_1} &=
\mathbb{F}_{\hat{f}_0}{}^{\hat{f}_1} \Gamma_{e\, \beta\,  \hat{B}}{}^{\hat{C}} \notag \\
&= \mathbb{F}_{\hat{f}_0}{}^{\hat{f}_1} \left({}^{\text{V}}\Gamma_{e\, \beta\,  \hat{B}}{}^{\hat{C}} +{}^{\text{S}}\Gamma_{e\, \beta\,  \hat{B}}{}^{\hat{C}} \right)\,,
\end{align}
with ${}^{\text{V}}\Gamma_e$ having the same apparent form as the single flavour solution prior to the introduction of the decipherer (see Ref.~\cite{Brill:1957fx}, note though our metric signature is $-2$ and the normalization for the $\gamma_0$ matrices is also slightly different from that paper)
\begin{align} \label{eq:GammaSol}
{}^{\text{V}}\Gamma_{e\, \alpha\, \hat{B}}{}^{\hat{A}} = -\frac{1}{2} g_{\beta \xi}\left[ \frac{\partial b_{\eta}{}^{\tilde{\delta}}}{\partial x^{\alpha}} a^{\xi}{}_{\tilde{\delta}}- \Gamma^{\xi}_{\eta \alpha} \right] \mathcal{B}^{\beta \eta}{}_{\hat{B}}{}^{\hat{A}}\,,
\end{align}
where $\Gamma^{\xi}_{\eta\alpha}$ are the Christoffel symbols, but the gamma matrices in $
\mathcal{B}_{\alpha\beta\, \hat{A}}{}^{\hat{C}} \equiv \frac{1}{2}\left\{\bar{\gamma}_{e\,\alpha \, \hat{D'}}{}^{\hat{C}}
\gamma_{e\, \beta \, \hat{A}}{}^{\hat{D'}}
-\left(\alpha \leftrightarrow \beta\right)\right\}$ now contain the decipherer in our context. On the other hand, the dynamics of the decipherer also introduce the additional direct contribution 
\bea \label{eq:NewGamma}
{}^{\text{S}}\Gamma_{e\, \beta\,  \hat{A}}{}^{\hat{E}} = -  \frac{\partial S_e^{-1}{}_{\tilde{B}}{}^{\hat{E}}}{\partial x^{\beta}}{S}_{e}{}^{\tilde{B}}{}_{\hat{A}}\,.
\eea
The simple flavour dependence \eqref{eq:SpinConnectionFactor} of the spin connection results in the flavoured Dirac-Weyl equation also being factorizable 
\begin{align} \label{eq:DiracWeylFull}
0&=\gamma^{\alpha}{}_{\hat{C} \, \hat{f}_2}{}^{\hat{B'} \, \hat{f}_1}\Big\{ \partial_{\alpha}\psi^{\hat{C} \,\hat{f}_2} +
 \Gamma_{\alpha}{}_{\hat{A}\, \hat{f}_0}{}^{\hat{C}\, \hat{f}_2}\psi^{\hat{A} \,\hat{f}_0} \Big\} \notag \\
 &=\frac{1}{3}
\gamma_e{}^{\alpha}{}_{\hat{C}}{}^{\hat{B'}}\Big\{ \epsilon_{\hat{A}}{}^{\hat{C}}\partial_{\alpha} +
\Gamma_{e\,\alpha}{}_{\hat{A}}{}^{\hat{C}}  \Big\} \mathbb{G}_{\hat{f}_0}{}^{\hat{f}_1}\psi^{\hat{A} \,\hat{f}_0} \,. 
\end{align}
Naively, one would want to apply a $\mathbb{G}^{-1}_{\hat{f}_2}{}^{\hat{f}_3}$ and decouple the different flavours. However, the determinant of $\mathbb{G}$ vanishes, so such an inversion is not possible, and the system of equations \eqref{eq:DiracWeylFull} is under-determined. This has interesting physical consequences which we will return to in Sec.~\ref{sec:Conc}. In the mean time, we provide a concrete example for how to compute the various quantities defined above in real spacetimes. 

\section{Worked example} 
We specialize to the Schwarzschild spacetime \cite{Schwarzschild:1916uq} for illustration, which due to the Birkhoff's theorem \cite{2005GReGr..37.2253J,1923rmp..book.....B} is also the solution relevant for regions surrounding, but not inside Earth. There exists a special orthonormal tetrad corresponding to the Kinnersley null tetrad \cite{Kinnersley:1969zza}, and the associated NP quantities can be looked up in standard literature \cite{ChandrasekharBook} ($*^{10}$), explicating the DEOM \eqref{eq:EOMCompT} into
\begin{align} \label{eq:EOMExterior}
4 r \frac{\partial C^{\hat{E}}_{\o}}{\partial r}&+2 \sqrt{2} \frac{\partial C^{\hat{E}}_{\iota}}{\partial \theta }-2 i \sqrt{2} \csc \theta  \frac{\partial C^{\hat{E}}_{\iota}}{\partial \phi }
=-\sqrt{2} \cot \theta  C^{\hat{E}}_{\iota}-4 C^{\hat{E}}_{\o}\,, \notag \\
2\frac{\Delta}{r} \frac{\partial C^{\hat{E}}_{\iota}}{\partial r}&-2 \sqrt{2} \frac{\partial C^{\hat{E}}_{\o}}{\partial \theta }-2 i \sqrt{2} \csc \theta  \frac{\partial C^{\hat{E}}_{\o}}{\partial \phi } =-2\left(1-\frac{M}{r}\right) C^{\hat{E}}_{\iota}+\sqrt{2} \cot \theta C^{\hat{E}}_{\o}\,. 
\end{align}
where $\Delta \equiv r(r-2M)$. We can make the decomposition $C^{\hat{E}}_{\o/\iota}=\mathring{C}^{\hat{E}}_{\o/\iota}+\delta^{\hat{E}}_{\o/\iota}$, where overhead circles indicate solutions in the limit of vanishing mass ($M\rightarrow 0$). The asymptotically flat spacetime boundary conditions can then be imposed simply as $\lim_{r\rightarrow \infty}\delta^{\hat{E}}_{\o/\iota}=0$. The $\mathring{C}$s are constants ($*^{8}$) in a Cartesian orthonormal tetrad, but not so in our tetrad adapted to spherical symmetry, so a tedious but straightforward transformation can be applied to acquire them ($*^{11}$), which subsequently serve as sources to the $\delta$s. 
When we are not too close to the event horizon, we can make a perturbative expansion $\delta^{\hat{E}}_{\o/\iota} = f^{\hat{E}}_{\o/\iota}(\theta,\phi)/r +\mathcal{O}\left(1/r^2\right)$, and Eqs.~\eqref{eq:EOMExterior} decouple at the leading order in $1/r$, into 
\begin{align}
&\cot\theta f^{\hat{E}}_{\iota}-2i\csc\theta\frac{\partial f^{\hat{E}}_{\iota}}{\partial \phi}+2 \frac{\partial f^{\hat{E}}_{\iota}}{\partial \theta} = 0\,,\notag \\
&\cot\theta f^{\hat{E}}_{\o}+2i\csc\theta\frac{\partial f^{\hat{E}}_{\o}}{\partial \phi}+2\frac{\partial f^{\hat{E}}_{\o}}{\partial \theta} = -\sqrt{2}M \mathring{C}^{\hat{E}}_{\iota}\,.
\end{align}
The solutions to these equations are
\begin{align} \label{eq:Extfo}
f^{\hat{E}}_{\iota} =& \frac{\mathfrak{g}^{\hat{E}}_{\iota}(\phi-i\tau)}{\sqrt{\sin\theta}}\,,\notag \\
f^{\hat{E}}_{\o} =& -\frac{M}{\sqrt{2\sin\theta}}\int_{1}^{\theta}d\theta'\sqrt{\sin\theta'}\mathring{C}^{\hat{E}}_{\iota}(\theta',\phi+i\tau-i\tau') +\frac{\mathfrak{g}^{\hat{E}}_{\o}(\phi+i\tau)}{\sqrt{\sin\theta}}\,,
\end{align}
where $\tau \equiv \ln(\cot(\theta/2))$ (similarly for $\tau'$ as a function of $\theta'$), and $\mathfrak{g}^{\hat{E}}_{\o/\iota}$ are single-variate functions with the entries in the trailing brackets their arguments (note the freedom in the lower integration limit is subsumed into $\mathfrak{g}^{\hat{E}}_{\o}$). They are solutions to the homogeneous parts of the equations, and are determined by initial conditions from the formation history of the central massive object.  

\begin{figure}[t!]
  \centering
\begin{overpic}[width=0.49\columnwidth]  {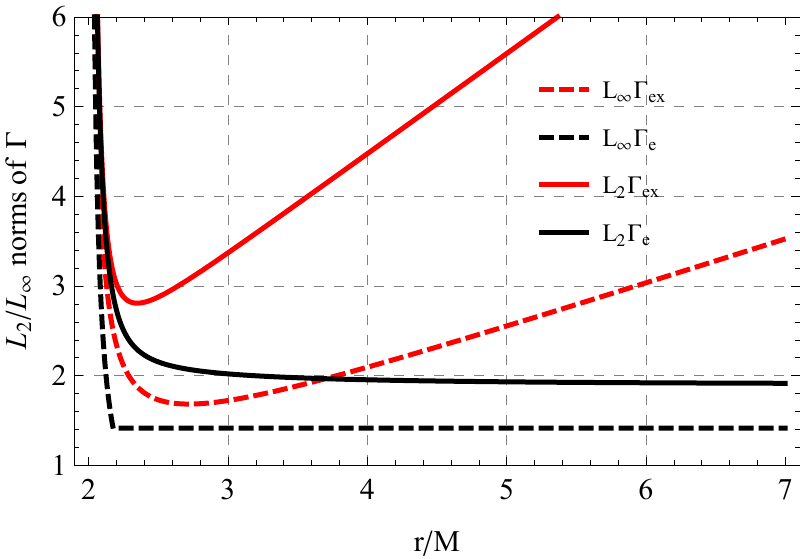}
\end{overpic}
  \caption{The $L_2$ and $L_{\infty}$ norms of $\Gamma_{\text{ex}}$ and $\Gamma_{e}$, evaluated as functions of $r$ at $\theta=\pi/2$ and $\phi=0$.  Note that although the plot extends down to the horizon to make all lines fit into the same figure, the underlying solution is only accurate to the $r^{-1}$ order.  
}
	\label{fig:GammaCompare}
\end{figure}

Concentrating on a specific example angular location such as $(\theta=\pi/2,\,\phi=0)$, we can numerically evaluate the integral in Eq.~\eqref{eq:Extfo} and its angular derivatives ($*^{12}$), yielding the spin connection via Eqs.~\eqref{eq:GammaSol} and \eqref{eq:NewGamma} ($*^{13}$). To demonstrate the changes brought about by the dynamic decipherer, we compare in Fig.~\ref{fig:GammaCompare} a reference $\Gamma_{\text{ex}}$ computed using $S_{e \, \hat{A}}{}^{\tilde{B}} \equiv \mathbb{1}_{\hat{A}}{}^{\tilde{B}}$ (to mimic the spin connection appearing in previous literature) with the full $\Gamma_e$. 
For this figure, we chose $\mathfrak{g}_{\iota}^{\hat{E}}=\mathfrak{g}_{\o}^{\hat{0}}=0$, and (to ensure regularity at the poles)
\bea
\mathfrak{g}_{\o}^{\hat{1}}=\frac{M}{2\sqrt{2}}\Big[(\lambda_n-\lambda_s)\tanh(i\tau)+(\lambda_n+\lambda_s)\Big]\,,
\eea
which straddles between $\lambda_n$ and $\lambda_s$ that are the integration results of Eq.~\eqref{eq:Extfo} when the upper limit $\theta$ is set for the north and south poles respectively. The $L_2$ norm plotted in the figure is defined as $\sqrt{\sum_{\alpha \, \hat{A}, \hat{B}}|\Gamma_{e/\text{ex}\,\alpha\, \hat{B}}{}^{\hat{A}}|^2}$, while the $L_{\infty}$ norm is $\text{max}_{\alpha \, \hat{A}, \hat{B}}|\Gamma_{e/\text{ex}\,\alpha\, \hat{B}}{}^{\hat{A}}|$, which can jump between different components and thus does not need to be smooth. 

The two spin connections exhibit visibly different behaviour at large $r$. Essentially, $\Gamma_{\text{ex}}$ contains within it a description of the variation of the spacetime spinor dyad (equivalently that of the Kinnersley tetrad), which is not necessarily due to gravity (see $*^{11}$ for details in the flat spacetime limit), just like the Christoffel symbols would not vanish even in the Minkowski spacetime if we use curvilinear coordinates. In any case, it is irrelevant to the particle spinor $\psi^{\hat{A}\, \hat{f}_0}$, and is neutralized by the decipherer dynamics introduced into $\Gamma_e$, which consequently constitutes a much cleaner representation of the effect of gravity on particle propagation.  

\section{Discussion}  \label{sec:Conc}
We conclude with a brief discussion on the physical implications of the flavoured Dirac-Weyl equation \eqref{eq:DiracWeylFull} being under-determined. Mathematically, this is because only a particular combination of the flavours ($\propto \bar{\mathcal{V}}^{\hat{f}_1}$) is visible to spacetime operators such as derivatives, and so the Dirac-Weyl equation can only guide this exposed part, leaving the remaining two flavour dimensions unconstrained. To see this, we can factor out an $\mathbb{F}$ from $\mathbb{G}$ using Eq.~\eqref{eq:GIdentities}, thus replacing $\psi^{A\, \hat{f}_0}$ in Eq.~\eqref{eq:DiracWeylFull} by its projection result $\mathbb{F}_{\hat{f}_0}{}^{\hat{f}_3}\psi^{A\, \hat{f}_0}$, meaning that the equation only cares about the component of $\psi^{A\, \hat{f}_0}$ along $\mathcal{V}^{\hat{f}_0}$. This is consistent with $\mathbb{G}_{\hat{f}_0}{}^{\hat{f}_1}=3\mathcal{V}_{\hat{f}_0}\bar{\mathcal{V}}^{\hat{f}_1}$ (Eq.~\ref{eq:GDecompose}), so only the said component contributes to a contraction with $\mathbb{G}$. However, due to the extra complex conjugation on $\bar{\mathcal{V}}^{\hat{f}_1}$, $\mathbb{G}$ itself is not generally a projection operator that preserves an already projected flavour vector through subsequent projections, even though it is always of rank one. So in the interest of physical clarity, although not mathematical necessity, it is desirable to choose the phase angles $\phi_{\mu}$ and $\phi_{\tau}$ ($*^8$) in such a way that $\mathbb{G}$ becomes some genre of a projection operator. 

In any case, the non-invertibility of $\mathbb{G}$ has the interesting implication that even without introducing any intrinsic masses, the Dirac-Weyl equation already allows for flavour oscillations. Because during propagation, $\psi^{A \, \hat{f}_0}$ is free to rotate within the two hidden flavour dimensions complementary to $\bar{\mathcal{V}}^{\hat{f}_1}$. For a more quantitative illustration, we can begin with any non-oscillating solution to Eq.~\eqref{eq:DiracWeylFull}, and then multiply it with any arbitrary holomorphic function of $\mathbb{F}$ (which we recall is not diagonal $*^{6}$) that also depends on the affine parameter $\tau$ along the spacetime trajectory of a neutrino, say $\exp(-i\Omega\tau \mathbb{F})_{\hat{f}_1}{}^{\hat{f}_0}=\mathbb{1}_{\hat{f}_1}{}^{\hat{f}_0}+(\exp(-i\Omega\tau)-1)\mathbb{F}_{\hat{f}_1}{}^{\hat{f}_0}$ ($\Omega$ being some constant), then the product is also a solution to Eq.~\eqref{eq:DiracWeylFull}, and its flavour evolution during propagation can be approximated by 
$i (d\psi^{\hat{A}\, \hat{f}_0}/d\tau) \sim \Omega\, \mathbb{F}_{\hat{f}_1}{}^{\hat{f}_0} \psi^{\hat{A}\,\hat{f}_1}$. This expression has an analogous form to the Hamiltonian depiction of flavour oscillations \cite{Nunokawa:2007qh} ($*^{14}$)(derived with the intrinsic mass explanation, but is more generally applicable if seen as a phenomenological model). In fact, fixing the PMNS parameter values to 
$\Delta m_{21} \ll \Delta m_{31}$, $\theta_{23}=\pi/4$, $\theta_{13}=\theta_{12}=\arcsin(1/\sqrt{3})$ and $\delta=3\pi/2$, the aforementioned Hamiltonian (not a projection operator in general) becomes exactly proportional to $\mathbb{F}$ with $\phi_{\mu}=\phi_{\tau}=\pi/2$, which happens to make $\mathbb{G}$ a ($\mathbb{Z}_2$-graded) projection operator satisfying $\mathbb{G}_{\hat{f}_2}{}^{\hat{f}_1}\mathbb{G}_{\hat{f}_0}{}^{\hat{f}_2}=-\mathbb{G}_{\hat{f}_0}{}^{\hat{f}_1}$. Aside from the special $\theta_{13}$ (burdened with the CP-violating $e^{-i\delta}\approx i$), these values are all close to experimental measurements \cite{Olive:2016xmw} ($*^{14}$), 
and thus may provide a tentative clue to answering perplexing questions such as why $\theta_{23}$ and $\delta$ are so nearly maximal, and why the inscrutable number $1/\sqrt{3}$ seems to lurk in the data (e.g., it is incorporated into the earlier phenomenological tribimaximal mixing scheme). 
{More precisely, we would like to impress upon the readers that the conditions for projection (including $\mathbb{Z}_2$-graded) operators are quite stringent (idempotency being a matrix equation), so if one starts from these conditions and go in the reverse direction to try and infer the mixing angles, they would arrive at essentially the same set of numbers (if $-2$ is picked for the right hand side of Eq.~\eqref{eq:ProjG}; slightly more freedom is allowed if $0$ is chosen). In this sense, these angles can be viewed as restrictive predictions of our phenomenological model, in order for it to achieve self-consistency. }
In contrast, the mixing angles are unconstrained in mass-based oscillation mechanisms, {so they are free parameters in those models, to be determined by fitting to data}. Our naive example is however not intended as a full-fledged model, and cannot explain all aspects of the recorded data. Nevertheless, the true flavour oscillation mechanism may well share its exploitation of $\mathbb{F}$ as a penalty-free (with respect to the Dirac-Weyl equation) channel for blending flavours. Therefore, the fact that some vestiges of $\mathbb{F}$ appear to remain detectable in experimental data should not be surprising. 

Physically, the flavour-projected $\mathbb{G}_{\hat{f}_0}{}^{\hat{f}_1}\psi^{\hat{A} \,\hat{f}_0}$ (really a fixed-flavour quantity due to the projection, despite the ${\hat{f}_1}$ index -- in an alternative flavour basis where $\bar{\mathcal{V}}^{\hat{f}_1}$ is along one base vector, the index ${\hat{f}_1}$ reduces to a fixed constant, leaving no free flavour indices) 
provides a handle that the Dirac-Weyl equation can grab onto, in order to ensure that the neutrinos travel on trajectories appropriate for massless spin $1/2$ particles being scattered by spacetime curvature (note that although a neutrino switches between flavours while propagating, it is still only one single particle, so it is appropriate that only one Dirac-Weyl equation should describe its propagation, instead of a separate equation for each flavour). How the internal structural details of the neutrinos evolve while they travel on said trajectories is on the other hand beyond the jurisdiction of the Dirac-Weyl equation, and requires additional beyond-SM physics. However, we note that only augmentations are required, and no modifications to the existing SM equations appear mandatory. 
While current efforts to explain neutrino flavour oscillations (see e.g.~\cite{1977PhLB...67..421M,2013arXiv1306.4669G,1980PThPh..64.1103Y,1980PhRvL..44..912M,1980PhRvD..22.2227S}) concentrate on the latter approach and are presently battling difficulties such as short baseline anomalies \cite{1995PhRvL..75.2650A,2001PhRvD..64k2007A,2013PhRvL.110p1801A,2016JPhCS.718c2008L,2016A&A...594A..13P} (additional light sterile neutrinos are proposed as potential solutions, but their existence contradicts results by collider, astrophysical, cosmological, as well as long baseline oscillation experiments \cite{Olive:2016xmw,2016JPhCS.718c2008L,2016A&A...594A..13P,2016JPhG...43c3001G,2016PhRvL.117g1801A})
and a lack of direct experimental support for the novel physics they employ \cite{KamLAND-Zen:2016pfg,2017arXiv170300570A,2016NuPhB.910....1B,2009PhRvL.102n1801N,2016JHEP...08..067A}, it is perhaps worthwhile to also explore the former (i.e., keep neutrinos massless and maximally parity violating, and simply devise new laws to govern their dynamics in those hidden flavour dimensions), and take advantage of a more straightforward reconciliation with existing tests of the SM. Experiments measuring neutrino masses directly, such as KATRIN \cite{2012NuPhS.229..146T}, should provide further clues as to whether this alternative line of attack is worth pursuing.  

{Finally, we note that even though the separation of the particle and spacetime spinors is presented in the context of general curved spacetimes in this work, the augmentations (inserted to describe the evolution of the particle spinors, prior to the application of the decipherer that subsequently determine the spacetime spinor) that cause the particle spinors to evolve with time in the flavour space do not rely on gravity, as the under-determinacy is also present in flat spacetimes. In other words, spacetime curvature causes the decipherer to become dynamic, but this has nothing to do with the (independent set of) dynamics within the particle spinor space (flavour space) responsible for the oscillations (note different particles should in general have different such dynamics, so non-neutrino Fermions do not need to oscillate). Regardless of whether the decipherer is static, it always contains a projection operator which keeps the Dirac-Weyl equation oblivious (thus will not object to, by failing to remain satisfied) to the extra flavour dynamics being inserted, provided that the additions do not alter the projection outcome (it is this provision that allowed us to construct the aforementioned phenomenological model and gave it some predictive power). It is therefore not necessary for gravity to be present to activate these flavour space dynamics, and in fact, the phenomenological model is intended for the Minkowski spacetime. Nevertheless, more complex interplays (beyond those already investigated by existing literature \cite{Cardall:1996cd,Cardall:1997bi,Fornengo:1997qu,Fornengo:1996ef,Mukhopadhyay:2007vca,Seitenzahl:2015mdy,Ahluwalia:1996ev,Konno:1998kq,Lambiase:2004qk,Lambiase:2005gt,Prasanna:2001ie}) between said flavour space dynamics and gravity are possible, and may well contribute to those residual differences between the phenomenological model's predictions and the experimental results. Furthermore, we make an observation that one feature separating long and short baseline oscillation experiments is that the longer ones all have the neutrinos tunnelling through Earth, and the spacetime curvature they experience would be of a purely Ricci nature (modelling the Earth as a uniform density ball of matter), as opposed to the Weyl-only curvatures experienced by neutrinos flying above the Earth's surface. We do not yet know how significant this difference is, but given the tension between the results emerging from long and short baseline experiments, it may deserve further investigation. }

\vspace{10mm}
{We thank an anonymous referee for their insightful comments that led to many of the detailed discussions in the final section of the paper}. The author is supported by the National Natural Science Foundation of China Grant No.~11503003 and 11633001, the Strategic Priority Research Program of the Chinese Academy of Sciences Grant No.~XDB23000000, and a Returned Overseas Chinese scholars Foundation grant. 


\appendix
\renewcommand{\theequation}{A.\arabic{equation}}
  \setcounter{equation}{0}  
  \section*{APPENDIX}  

\section{Additional details} \label{sec:App}
{\bf *1-} The spacetime spinors represented in the spinor basis frame (dyad) $(\o^{\tilde{A}},\iota^{\tilde{A}})$ can be seen as exotic representations of the spacetime vectors expressed in a Newman-Penrose null tetrad $\{\ell,n,m,\bar{m}\}$ given by 
\bea \label{eq:SimpleMapping}
\ell^{\alpha} &=& \gamma_{0\,\tilde{A}\tilde{A'}}{}^{\alpha} \o^{\tilde{A}}\bar{\o}^{\tilde{A'}}\,, \quad n^{\alpha} = \gamma_{0\,\tilde{A}\tilde{A'}}{}^{\alpha} \iota^{\tilde{A}}\bar\iota^{\tilde{A'}}\,, \notag \\ 
m^{\alpha} &=& \gamma_{0\,\tilde{A}\tilde{A'}}{}^{\alpha}\o^{\tilde{A}}\bar\iota^{\tilde{A'}}\,, 
\quad \bar m^{\alpha} = \gamma_{0\,\tilde{A}\tilde{A'}}{}^{\alpha} \iota^{\tilde{A}}\bar{\o}^{\tilde{A'}}\,, 
\eea
which is equivalent to one particular orthonormal frame $\{T, N, E_2, E_3\}$ via
\begin{align}\label{eq:Orthonormal}
T^{\alpha}&=\frac{1}{\sqrt{2}}\left(\ell^{\alpha}+n^{\alpha}\right)\,, \notag \\
N^{\alpha}&=\frac{1}{\sqrt{2}}\left(\ell^{\alpha}-n^{\alpha}\right)\,, \notag \\ 
E^{\alpha}_2 &= \frac{1}{\sqrt{2}}\left(m^{\alpha}+\bar{m}^{\alpha}\right)\,, 
\notag \\
E^{\alpha}_3 &= \frac{i}{\sqrt{2}}\left(m^{\alpha}-\bar{m}^{\alpha}\right)\,.
\end{align}
The $\gamma_0$s appearing in Eq.~\eqref{eq:SimpleMapping} are by construction constant matrices such as the Pauli matrices joined by the identity matrix (apart from a common normalization factor, see Ref.~\cite{Penrose1992} Eq.~3.1.49) 	
\begin{align} \label{eq:FlatSoldering}
&\gamma_{0\, 0}{}^{\tilde{A}\tilde{A'}} = \frac{1}{\sqrt{2}}\bma 1 & 0 \\ 0 & 1 \ema = \gamma_{0\, \tilde{A}\tilde{A'}}{}^0 \,, \notag \\
&\gamma_{0\, 1}{}^{\tilde{A}\tilde{A'}} = \frac{1}{\sqrt{2}}\bma 0 & 1 \\ 1 & 0 \ema= \gamma_{0\, \tilde{A}\tilde{A'}}{}^1 \,, \notag \\
&\gamma_{0\, 2}{}^{\tilde{A}\tilde{A'}} = \frac{1}{\sqrt{2}}\bma 0 & i \\ -i & 0 \ema = -\gamma_{0\, \tilde{A}\tilde{A'}}{}^2 \,, \notag \\
&\gamma_{0\, 3}{}^{\tilde{A}\tilde{A'}} = \frac{1}{\sqrt{2}}\bma 1 & 0 \\ 0 & -1 \ema= \gamma_{0\, \tilde{A}\tilde{A'}}{}^3 \,.
\end{align}
\newline\newline
{\bf *2-} With $\eta_{\tilde{\alpha}\tilde{\beta}} = \tilde{g}_{\alpha\beta} \equiv a^{\xi}{}_{\tilde{\alpha}} a{}^{\zeta}{}_{\tilde{\beta}} g_{\xi \zeta}$ being the Minkowski metric under the orthonormal frame, we have 
\begin{align} \label{eq:SGauge}
&\bar{\gamma}_{\alpha \, \hat{A'}}{}^{\hat{D}}\gamma_{\beta\,\hat{D}}{}^{\hat{C'}}+\bar{\gamma}_{\beta\,\hat{A'}}{}^{\hat{D}}\gamma_{\alpha\,\hat{D}}{}^{\hat{C'}} 
\notag \\
=&S^{-1}{}_{\tilde{B}}{}^{\hat{D}}
\bar{\tilde{\gamma}}_{0\,\alpha\, \tilde{B'}}{}^{\tilde{B}} \bar{S}{}^{\tilde{B'}}{}_{\hat{A'}}
\bar{S}^{-1}{}_{\tilde{E'}}{}^{\hat{C'}}\tilde{\gamma}_{0\, \beta \, \tilde{F}}{}^{\tilde{E'}}S{}^{\tilde{F}}{}_{\hat{D}}
+\alpha \leftrightarrow \beta \notag \\
=&
\bar{\tilde{\gamma}}_{0\,\alpha\, \tilde{B'}}{}^{\tilde{B}} \bar{S}{}^{\tilde{B'}}{}_{\hat{A'}}
\bar{S}^{-1}{}_{\tilde{E'}}{}^{\hat{C'}}\tilde{\gamma}_{0\, \beta \, \tilde{F}}{}^{\tilde{E'}}\epsilon_{\tilde{B}}{}^{\tilde{F}}
+\alpha \leftrightarrow \beta \notag \\
=&
\bar{S}{}^{\tilde{B'}}{}_{\hat{A'}}\bar{\tilde{\gamma}}_{0\,\alpha\, \tilde{B'}}{}^{\tilde{B}} 
\tilde{\gamma}_{0\, \beta \, \tilde{B}}{}^{\tilde{E'}}\bar{S}^{-1}{}_{\tilde{E'}}{}^{\hat{C'}}
+\alpha \leftrightarrow \beta \notag \\
=&
\bar{S}{}^{\tilde{B'}}{}_{\hat{A'}}b{}_{\alpha}{}^{\tilde{\xi}}\bar{{\gamma}}_{0\,\tilde{\xi}\, \tilde{B'}}{}^{\tilde{B}} 
b{}_{\beta}{}^{\tilde{\zeta}}{\gamma}_{0\, \tilde{\zeta} \, \tilde{B}}{}^{\tilde{E'}}\bar{S}^{-1}{}_{\tilde{E'}}{}^{\hat{C'}}
+ \tilde{\xi} \leftrightarrow \tilde{\zeta} \notag \\
=&
-\bar{S}{}^{\tilde{B'}}{}_{\hat{A'}}b{}_{\alpha}{}^{\tilde{\xi}}b{}_{\beta}{}^{\tilde{\zeta}}
\tilde{g}_{\tilde{\xi}\tilde{\zeta}}\epsilon_{\tilde{B'}}{}^{\tilde{E'}}
\bar{S}^{-1}{}_{\tilde{E'}}{}^{\hat{C'}}
\notag \\
=&
-b{}_{\alpha}{}^{\tilde{\xi}}b{}_{\beta}{}^{\tilde{\zeta}}
\tilde{g}_{\tilde{\xi}\tilde{\zeta}}
\bar{S}_{\hat{A'}}{}^{\tilde{B'}}\bar{S}^{-1}{}_{\tilde{B'}}{}^{\hat{C'}}
\notag \\
=&-g_{\alpha \beta} \epsilon_{\hat{A'}}{}^{\hat{C'}} \,.
\end{align}
\newline\newline
{\bf *3-} The decipherer translates between the particle and spacetime spinor dyads. The latter has been fixed by the tetrad choice, but the former can be altered, resulting in a corresponding change in the decipherer $S$. In other words, there is a global gauge freedom (we shall term it the representational freedom) in $S$ on top of its satisfying the decipherer equation of motion. Such a gauge transformation can be written out as 
\bea \label{eq:GlobalGauge}
\gamma_{\alpha \, \hat{A}}{}^{\hat{A'}} \rightarrow \bar{G}^{-1}{}_{\hat{B'}}{}^{\hat{A'}}\gamma_{\alpha \, \hat{B}}{}^{\hat{B'}}G_{\hat{A}}{}^{\hat{B}}\,,
\eea
where $G$ (both indices are hatted, as it is a transformation within the particle spinor space) can obviously be subsumed into $S$ so the gauge transformation can alternatively be stated as a change in the decipherer, which does not alter the observable quantities.   
More precisely, given the Dirac-Weyl equation 
\bea
\gamma_{\hat{A}\hat{A'}}{}^{\alpha}\left(\epsilon_{\hat{B}}{}^{\hat{A}}\partial_{\alpha} + \Gamma_{\alpha\, \hat{B}}{}^{\hat{A}} \right) {\psi}^{\hat{B}} = 0\,,
\eea
the transformation \eqref{eq:GlobalGauge} induces a corresponding alteration in $\Gamma$
\bea \label{eq:GlobalGauge2}
\Gamma_{\alpha \, \hat{A}}{}^{\hat{B}} \rightarrow {G}^{-1}{}_{\hat{C}}{}^{\hat{B}}\Gamma_{\alpha \, \hat{D}}{}^{\hat{C}}G_{\hat{A}}{}^{\hat{D}}\,,
\eea
which we can verify to satisfy the metricity condition, explicitly 
\begin{align} 
&\bar{G}^{-1}{}_{\hat{B'}}{}^{\hat{A'}} G_{\hat{A}}{}^{\hat{B}}
\partial_{\beta}\gamma_{\alpha\, \hat{B}}{}^{\hat{B'}} 
- \bar{G}^{-1}{}_{\hat{B'}}{}^{\hat{A'}}G_{\hat{A}}{}^{\hat{B}}
\Gamma^{\xi}_{\alpha \beta} \gamma_{\xi\, \hat{B}}{}^{\hat{B'}} + 
\bar{G}^{-1}{}_{\hat{F}}{}^{\hat{A'}}\bar{\Gamma}_{\beta\, \hat{E'}}{}^{\hat{F}}\bar{G}_{\hat{B'}}{}^{\hat{E'}}
\bar{G}^{-1}{}_{\hat{D'}}{}^{\hat{B'}}\gamma_{\alpha\, \hat{C}}{}^{\hat{D'}}G_{\hat{A}}{}^{\hat{C}} 
\notag \\
&-{G}^{-1}{}_{\hat{F}}{}^{\hat{B}}\Gamma_{\beta\, \hat{E}}{}^{\hat{F}}G_{\hat{A}}{}^{\hat{E}}
\bar{G}^{-1}{}_{\hat{D'}}{}^{\hat{A'}}\gamma_{\alpha\, \hat{C}}{}^{\hat{D'}}G_{\hat{B}}{}^{\hat{C}} 
\notag \\
=&\bar{G}^{-1}{}_{\hat{B'}}{}^{\hat{A'}} G_{\hat{A}}{}^{\hat{B}}
\Big[
\partial_{\beta}\gamma_{\alpha\, \hat{B}}{}^{\hat{B'}}
- \Gamma^{\xi}_{\alpha \beta} \gamma_{\xi\, \hat{B}}{}^{\hat{B'}} + 
\bar{\Gamma}_{\beta\, \hat{D'}}{}^{\hat{B'}}\gamma_{\alpha\, \hat{B}}{}^{\hat{D'}} 
-\Gamma_{\beta\, \hat{B}}{}^{\hat{C}}
\gamma_{\alpha\, \hat{C}}{}^{\hat{B'}} \Big] =0\,.
\end{align}
The Dirac-Weyl equation then remains satisfied, provided that $\psi$ undergoes a corresponding transformation
\bea
\psi^{\hat{A}} \rightarrow G^{-1}{}_{\hat{B}}{}^{\hat{A}}\psi^{\hat{B}}\,,
\eea
explicitly 
\begin{align}
\bar{G}^{-1}{}_{\hat{C'}}{}^{\hat{A'}}\gamma_{\hat{C}}{}^{\hat{C'}\, \alpha}G_{\hat{A}}{}^{\hat{C}} \left(\epsilon_{\hat{B}}{}^{\hat{A}}\partial_{\alpha} + {G}^{-1}{}_{\hat{D}}{}^{\hat{A}}\Gamma_{\alpha \, \hat{C}}{}^{\hat{D}}G_{\hat{B}}{}^{\hat{C}}\right) G^{-1}{}_{\hat{E}}{}^{\hat{B}}{\psi}^{\hat{E}}=\bar{G}^{-1}{}_{\hat{C'}}{}^{\hat{A'}}\left[\gamma_{\hat{C}}{}^{\hat{C'}\, \alpha}\left(  \epsilon_{\hat{E}}{}^{\hat{C}} \partial_{\alpha}+ \Gamma_{\alpha \, \hat{E}}{}^{\hat{C}}\right){\psi}^{\hat{E}}\right] =0 \,.
\end{align} 
The physical quantities such as the current density remain unchanged, explicitly 
\begin{align}
\bar{\psi}^{\hat{B'}}\bar{G}^{-1}{}_{\hat{B'}}{}^{\hat{A'}}
\bar{G}{}_{\hat{A'}}{}^{\hat{C'}}\gamma_{\hat{C}\hat{C'}}{}^{\alpha}G_{\hat{A}}{}^{\hat{C}}G^{-1}{}_{\hat{B}}{}^{\hat{A}}{\psi}^{\hat{B}}
=\bar{\psi}^{\hat{B'}}\gamma_{\hat{B}\hat{B'}}{}^{\alpha}{\psi}^{\hat{B}}\,,
\end{align}
where we have used the fact that the contravariant and covariant quantities transform with inverted matrices (one with $G$, the other with $G^{-1}$), in order for their contraction to remain invariant. 
\newline\newline
{\bf *4-} The freely propagating massless field equation implies 
\begin{align} \label{eq:DeriveEOM}
0=& \bar{S}^{-1}_{\tilde{A'}}{}^{\hat{E'}}\nabla^{\tilde{A}\tilde{A'}}\left(S^{-1}_{\tilde{A}}{}^{\hat{E}}\psi_{\hat{E}}\right)\notag \\
=& \bar{S}^{-1}_{\tilde{A'}}{}^{\hat{E'}}S^{-1}_{\tilde{A}}{}^{\hat{E}}\nabla^{\tilde{A}\tilde{A'}}\psi_{\hat{E}}
 +\bar{S}^{-1}_{\tilde{A'}}{}^{\hat{E'}}\psi_{\hat{E}}\nabla^{\tilde{A}\tilde{A'}}S^{-1}_{\tilde{A}}{}^{\hat{E}}\notag \\
 =& \nabla^{\hat{E}\hat{E'}}\psi_{\hat{E}}
 +\bar{S}^{-1}_{\tilde{A'}}{}^{\hat{E'}}\psi_{\hat{E}}\nabla^{\tilde{A}\tilde{A'}}S^{-1}_{\tilde{A}}{}^{\hat{E}}\,,
\end{align}
and we know that the first term of the last line vanishes since it is just the Dirac-Weyl equation in its traditional form. Because Eq.~\eqref{eq:DeriveEOM} has to be satisfied for all possible neutrino fields $\psi$, we must have that the derivative in the second term vanishes. 
\newline\newline
{\bf *5-} Given the metric for a spacetime, we begin by specifying any particular Newman-Penrose null tetrad (thus fixing $a^{\beta}{}_{\tilde{\alpha}}$). It then defines a corresponding spin-frame via Eq.~\eqref{eq:SimpleMapping}, or
\footnote{\label{ft:ph} Writing $\iota^A = \iota^1(\iota^0/\iota^1,1)$ and $\o^A = \o^1(\o^0/\o^1,1)$, then the dyad satisfying the complex normalization equation \eqref{eq:Normalization} fixes $|\iota^1||\o^1|$ and $\arg \iota^1 + \arg \o^1$. Specializing to a particular null tetrad then further fixes the three complex Lorentz freedom transforming between orthonormal tetrads. Specifically, $\iota^0/\iota^1$ (direction of $n$ on the anti-celestial sphere in a complex stereographic projection representation), $\o^0/\o^1$ (direction of $l$), $|\iota^1|/|\o^1|$ (overall multiplicative amplitude of $l$ or equivalently $n$), and $\arg \iota^1 - \arg \o^1$ (orientation of $E_2$ or equivalently $E_3$ that's orthogonal to it) are fixed.} 
\begin{align} \label{eq:DyadMap}
\left[\tilde{\gamma}_{0\, \beta}{}^{\tilde{A}\tilde{A'}}\right]\left(\ell^{\beta},n^{\beta},m^{\beta}\right) 
&=\left[\gamma_{0\, \tilde{\beta}}{}^{\tilde{A}\tilde{A'}}\right]\left(\ell^{\tilde\beta},n^{\tilde\beta},m^{\tilde\beta}\right) \notag \\
&= \left(\o^{\tilde{A}} \bar{\o}^{\tilde{A'}}, \iota^{\tilde{A}} \bar{\iota}^{\tilde{A'}}, \o^{\tilde{A}} \bar{\iota}^{\tilde{A'}} \right)\,, 
\end{align}
where a spin-frame is a spinor dyad $(\o^{\tilde{A}},\iota^{\tilde{A}})$ that satisfies the normalization 
\bea \label{eq:Normalization}
\o_{\tilde{A}}\iota^{\tilde{A}}=1=-\iota_{\tilde{A}}\o^{\tilde{A}}\,,
\eea
and is similar to the concept of an ``orthonormal tetrad" with tensors. 
The antisymmetric bispinors that are used to lower spinor indices can then be written out as (we do not place a bar over $\epsilon_{\tilde{A'}\tilde{B'}}$, as per convention)
\bea \label{eq:Symp}
\epsilon_{\tilde{A}\tilde{B}} = {\o}_{\tilde{A}}{\iota}_{\tilde{B}}-{\iota}_{\tilde{A}}{\o}_{\tilde{B}}\,, \quad 
\epsilon_{\tilde{A'}\tilde{B'}} = \bar{\o}_{\tilde{A'}}\bar{\iota}_{\tilde{B'}}-\bar{\iota}_{\tilde{A'}}\bar{\o}_{\tilde{B'}}\,.
\eea
The component form of Eq.~\eqref{eq:Orthonormal} is (decomposed into the orthonormal tetrad $\{T,N,E_2,E_3\}$)
\bea
\ell^{\tilde{\beta}} = \frac{1}{\sqrt{2}}\bma 1 \\ 1 \\ 0 \\0 \ema\,,  \quad 
n^{\tilde{\beta}} = \frac{1}{\sqrt{2}}\bma 1 \\ -1 \\ 0 \\0 \ema\,,  \notag \\
m^{\tilde{\beta}} = \frac{1}{\sqrt{2}}\bma 0 \\ 0 \\ 1 \\ -i \ema\,,  \quad 
\bar{m}^{\tilde{\beta}} = \frac{1}{\sqrt{2}}\bma 0 \\ 0 \\ 1 \\ i \ema\,,
\eea
which gives via Eq.~\eqref{eq:DyadMap} the component form of the spinor dyad as 
\bea \label{eq:dyadExp}
\o^{\tilde{A}}  =\frac{1}{\sqrt{2}}\bma  1 \\ 1 \ema\,, &\quad& 
\iota^{\tilde{A}}  =\frac{i}{\sqrt{2}}\bma  1 \\ -1 \ema\,, \notag \\
\o_{\tilde{A}} = \frac{i}{\sqrt{2}}\bma -1 \,,& 1 \ema \,, &\quad& 
\iota_{\tilde{A}} = \frac{1}{\sqrt{2}}\bma -1 \,,&  -1 \ema \,. 
\eea

The covariant derivatives in the DEOM acting on the spinors are more conveniently expressed in the Newman-Penrose formalism, whose intrinsic derivatives along the tetrad basis vectors are related to spinor derivatives by (Ref.~\cite{Penrose1992} Eq.~4.5.23)
\begin{align} \label{eq:NPDeriv}
D&=\bar{D}=\o^{\tilde{A}}\bar\o^{\tilde{A'}}{\nabla}_{\tilde{A}\tilde{A'}}=\ell^{\alpha}\nabla_{\alpha}\,, \notag \\
\delta&=\bar{\delta'}=\o^{\tilde{A}}\bar\iota^{\tilde{A'}}{\nabla}_{\tilde{A}\tilde{A'}}=m^{\alpha}\nabla_{\alpha}\,, \notag \\
D'&=\bar{D'}=\iota^{\tilde{A}}\bar\iota^{\tilde{A'}}{\nabla}_{\tilde{A}\tilde{A'}}=n^{\alpha}\nabla_{\alpha}\,, \notag \\
\delta'&=\bar{\delta}=\iota^{\tilde{A}}\bar\o^{\tilde{A'}}{\nabla}_{\tilde{A}\tilde{A'}}=\bar{m}^{\alpha}\nabla_{\alpha}\,. 
\end{align}
Making the decomposition of $S$ into the dyad $(\o,\iota)$ as in the main text, its covariant derivatives then break down by the additive and product rules (4.4.11 and 4.4.12 of Ref.~\cite{Penrose1992}) into simple partial derivatives on the coefficient scalars $C(x)$, as well as covariant derivatives on the dyad basis. The latter evaluates into the spin coefficients as in Eqs.~4.5.26 and 4.5.27 of Ref.~\cite{Penrose1992} (see also 4.5.29 for notation) 
\begin{align} \label{eq:dyadSpinCoe}
D\o^{\tilde{A}} &= \epsilon \o^{\tilde{A}} - \kappa \iota^{\tilde{A}}\,,
\quad 
D\iota^{\tilde{A}} = -\epsilon \iota^{\tilde{A}} + \pi\o^{\tilde{A}}\,, \notag \\
\delta'\o^{\tilde{A}} &= \alpha \o^{\tilde{A}} - \rho \iota^{\tilde{A}}\,,
\quad 
\delta'\iota^{\tilde{A}} = -\alpha \iota^{\tilde{A}} + \lambda\o^{\tilde{A}}\,, \notag \\
\delta\o^{\tilde{A}} &= \beta \o^{\tilde{A}} - \sigma \iota^{\tilde{A}}\,,
\quad 
\delta\iota^{\tilde{A}} = -\beta\iota^{\tilde{A}} + \mu\o^{\tilde{A}}\,, \notag \\
D'\o^{\tilde{A}} &= \gamma \o^{\tilde{A}} - \tau \iota^{\tilde{A}}\,,
\quad 
D'\iota^{\tilde{A}} = -\gamma \iota^{\tilde{A}} + \nu\o^{\tilde{A}}\,. 
\end{align}
Using the product rule and noting that the covariant derivatives of $\epsilon_{\tilde{A}\tilde{B}}$ and  $\epsilon_{\tilde{A'}\tilde{B'}}$ vanish, the derivatives of the lower-indexed dyad basis are trivially obtained by shifting the index location in these expressions. Their complex conjugations when combined with 
Eq.~\eqref{eq:NPDeriv} also provide the derivatives of $\bar\o^{\tilde{A'}}$ and $\bar\iota^{\tilde{A'}}$. Also notice that the spin coefficients in Eq.~\eqref{eq:dyadSpinCoe} that are components of the spin connection (note, this spin connection is for the spacetime spinors, different from the one acting on particle spinors that appear in the Dirac-Weyl equation) do not need to vanish, even in flat spacetimes, for spinor dyads (and correspondingly the null tetrads) that are not parallelly transported. 

We now turn to the DEOM, which becomes
\begin{align}
0=&\nabla^{\tilde{A}\tilde{A'}}S^{-1}_{\tilde{A}}{}^{\hat{E}} \notag \\
=& C^{\hat{E}}_{\o} \nabla^{\tilde{A}\tilde{A'}}\epsilon_{\tilde{B}\tilde{A}} \o^{\tilde{B}} + C^{\hat{E}}_{\iota} \nabla^{\tilde{A}\tilde{A'}}\epsilon_{\tilde{B}\tilde{A}} \iota^{\tilde{B}} + \o_{\tilde{A}} \nabla^{\tilde{A}\tilde{A'}}C^{\hat{E}}_{\o} + \iota_{\tilde{A}} \nabla^{\tilde{A}\tilde{A'}}C^{\hat{E}}_{\iota} \,.
\end{align}
Commuting $\epsilon_{\tilde{A}\tilde{B}}$ outside of the derivative and break it down using Eqs.~\eqref{eq:Symp}, before contracting with $\bar\o_{\tilde{A'}}$ and $\bar\iota_{\tilde{A'}}$ respectively, we obtain after applying Eq.~\eqref{eq:NPDeriv} the equations
\begin{align}
0=&C^{\hat{E}}_{\o}(\o_{\tilde{B}}\delta'-\iota_{\tilde{B}} D)\o^{\tilde{B}}+C^{\hat{E}}_{\iota}(\o_{\tilde{B}}\delta'-\iota_{\tilde{B}} D)\iota^{\tilde{B}} +DC^{\hat{E}}_{\o}+\delta' C^{\hat{E}}_{\iota} \,, \notag \\
0=&C^{\hat{E}}_{\o}(\o_{\tilde{B}} D'-\iota_{\tilde{B}} \delta)\o^{\tilde{B}}+C^{\hat{E}}_{\iota}(\o_{\tilde{B}} D'-\iota_{\tilde{B}} \delta)\iota^{\tilde{B}} +\delta C^{\hat{E}}_{\o}+D' C^{\hat{E}}_{\iota} \,.
\end{align}
Finally, Eq.~\eqref{eq:dyadSpinCoe} together with the normalization rule \eqref{eq:Normalization} give us 
\begin{align} \label{eq:EOMComp}
(\epsilon-\rho)C^{\hat{E}}_{\o}+(\pi-\alpha)C^{\hat{E}}_{\iota}+DC^{\hat{E}}_{\o}+\delta'C^{\hat{E}}_{\iota}&=0\,,\notag \\
(\beta-\tau)C^{\hat{E}}_{\o}+(\mu-\gamma)C^{\hat{E}}_{\iota}+\delta C^{\hat{E}}_{\o}+D'C^{\hat{E}}_{\iota}&=0\,,
\end{align}
where the differential operators appearing above are simply the far right hand side of Eq.~\eqref{eq:NPDeriv}, but with $\nabla$ replaced by $\partial$. 
\newline\newline
{\bf *6-} 
We first note that with the inverse $S^{-1}$ as defined in the main text, the derivation in *2 carries over essentially unchanged, leading to 
\begin{align} \label{eq:CliffordFull}
\Big[\bar{\gamma}_{\alpha\,\hat{D'}}{}^{\hat{C}}{}_{\hat{f}_0}{}^{\hat{f}_2}\Big] 
\Big[{\gamma}_{\beta\,\hat{A}}{}^{\hat{D'}}{}_{\hat{f}_1}{}^{\hat{f}_0}\Big]
+&\Big[\bar{\gamma}_{\beta\,\hat{D'}}{}^{\hat{C}}{}_{\hat{f}_0}{}^{\hat{f}_2}\Big]\Big[{\gamma}_{\alpha \, \hat{A}}{}^{\hat{D'}}{}_{\hat{f}_1}{}^{\hat{f}_0}\Big]
=-g_{\alpha \beta} \epsilon_{\hat{A}}{}^{\hat{C}} \mathbb{F}_{\hat{f}_1}{}^{\hat{f}_2}\,.
\end{align}
The spinor algebra is thus not spoiled, but we do not know \emph{a priori} what $\mathbb{F}$ should be, given the mystery shrouding generational physics. Nevertheless, we know it cannot be a simple Dirac delta. Writing out the flavour components as 
\bea \label{eq:SComp}
S_{}^{\tilde{B}}{}_{\hat{C}\, \hat{f}_0}= \frac{1}{\sqrt{3}}\bma 
S_{e}{}^{\tilde{B}}{}_{\hat{C}}\,, &
S_{\mu}{}^{\tilde{B}}{}_{\hat{C}}\,, & 
S_{\tau}{}^{\tilde{B}}{}_{\hat{C}}
\ema\,,
\eea
then the diagonal entries (in flavour) in the left-inverse and the whole of the right-inverse conditions demand that 
\begin{align} \label{eq:SCompInv}
&{S}^{-1}_{\tilde{A}}{}^{\hat{C} \, \hat{f}_0} = \frac{1}{\sqrt{3}}\bma S_{e}^{-1}{}_{\tilde{A}}{}^{\hat{C}} \\
S_{\mu}^{-1}{}_{\tilde{A}}{}^{\hat{C}} \\
S_{\tau}^{-1}{}_{\tilde{A}}{}^{\hat{C}} 
\ema\,, 
\end{align}
If $\mathbb{F}_{\hat{f}_1}{}^{\hat{f}_2}=\delta_{\hat{f}_1}{}^{\hat{f}_2}$ however, then we need $\text{tr}(\mathbb{F})=3$, which is impossible. We cannot even have $\mathbb{F}_{\hat{f}_1}{}^{\hat{f}_2}=\lambda \delta_{\hat{f}_1}{}^{\hat{f}_2}$ for some constant $\lambda$, since the vanishing of all off-diagonal elements in the left-inverse condition demands that
at least some of $\det(S_e)$, $\det(S_{\mu})$ or $\det(S_{\tau})$ vanish, making them non-invertible. Retaining invertibility, we then have the following conditions
\begin{align} \label{eq:Invertibility}
S_{\mu}{}^{\tilde{B}}{}_{\hat{A}} = \mathcal{F}_{\mu} \, S_{e}{}^{\tilde{B}}{}_{\hat{A}} \,, \quad 
S_{\tau}{}^{\tilde{B}}{}_{\hat{A}} = \mathcal{F}_{\tau} \, S_{e}{}^{\tilde{B}}{}_{\hat{A}} \,,  
\end{align}
that must be satisfied for some scalar fields $\mathcal{F}_{\mu}$ and $\mathcal{F}_{\tau}$. The $S_{e}$ is unconstrained, aside from the determinants $\det S_e$, $\det S_{\mu} = \mathcal{F}_{\mu}^2\det S_e$ and $\det S_{\tau} = \mathcal{F}_{\tau}^2\det S_e$
having to be nonvanishing. The corresponding $\mathbb{F}$ is 
\bea \label{eq:FDef}
\quad \quad \mathbb{F}_{\hat{f}_0}{}^{\hat{f}_1}= 
\frac{1}{3}\bma
1 & \mathcal{F}_{\mu} & \mathcal{F}_{\tau} \\
1/\mathcal{F}_{\mu} & 1 & \mathcal{F}_{\tau}/\mathcal{F}_{\mu}\\
1/\mathcal{F}_{\tau} &
\mathcal{F}_{\mu}/\mathcal{F}_{\tau}& 1
\ema \,,
\eea
which is a projection operator ($\mathbb{F}_{\hat{f}_2}{}^{\hat{f}_1}\mathbb{F}_{\hat{f}_0}{}^{\hat{f}_2}=\mathbb{F}_{\hat{f}_0}{}^{\hat{f}_1}$) that acts on flavour vectors in the following way
\begin{align} \label{eq:EigVec}
\mathbb{F}_{\hat{f}_0}{}^{\hat{f}_1}\xi^{\hat{f}_0} =\left({\mathcal{V}}_{\hat{f}_0}\xi^{\hat{f}_0} \right) \mathcal{V}^{\hat{f}_1}\,, \quad 
\mathcal{V}^{\hat{f}_1}=
\frac{1}{\sqrt{3}}\bma
1 \\
\mathcal{F}^{-1}_{\mu}\\
\mathcal{F}^{-1}_{\tau}
\ema
\,.
\end{align}
On the other hand, the explicit form for the flavour factor in the gamma matrices is
\begin{align} \label{eq:GDef}
\mathbb{G}_{\hat{f}_0}{}^{\hat{f}_1} = 
\bma
1 & \mathcal{F}_{\mu}/\mathcal{F}_{e} & 	\mathcal{F}_{\tau}/\mathcal{F}_{e} \\
\bar{\mathcal{F}}_{e}/\bar{\mathcal{F}}_{\mu} & 
(\mathcal{F}_{\mu}\bar{\mathcal{F}}_{e})/(\mathcal{F}_{e}\bar{\mathcal{F}}_{\mu}) &  
(\mathcal{F}_{\tau}\bar{\mathcal{F}}_{e})/(\mathcal{F}_{e}\bar{\mathcal{F}}_{\mu}) 
\\
\bar{\mathcal{F}}_{e}/\bar{\mathcal{F}}_{\tau} & 
(\mathcal{F}_{\mu}\bar{\mathcal{F}}_{e})/(\mathcal{F}_{e}\bar{\mathcal{F}}_{\tau}) & 
(\mathcal{F}_{\tau}\bar{\mathcal{F}}_{e})/(\mathcal{F}_{e}\bar{\mathcal{F}}_{\tau}) 
\ema\,,
\end{align}
which can be checked explicitly to satisfy the useful identities 
\bea  \label{eq:FlavourIdentity}
\bar{\mathbb{F}}{}_{\hat{f}_3}{}^{\hat{f}_1}\mathbb{G}_{\hat{f}_2}{}^{\hat{f}_3} = \mathbb{G}_{\hat{f}_2}{}^{\hat{f}_1}\,,
\quad 
\mathbb{G}_{\hat{f}_2}{}^{\hat{f}_3}\mathbb{F}{}_{\hat{f}_0}{}^{\hat{f}_2} = \mathbb{G}_{\hat{f}_0}{}^{\hat{f}_3} \,. 
\eea
\newline\newline
{\bf *7-} The derivations in (*4) remain unchanged after we augment the hatted spinor indices with flavour indices, because the spacetime indices that the derivative operators interact with are unaltered, so the evolution equation for the decipherer is just 
\bea \label{eq:deciphereEqFull}
\nabla^{\tilde{A}\tilde{A'}}S^{-1}_{\tilde{A}}{}^{ \hat{E} \, \hat{f}_0} =0\,,
\eea
where each flavour can be evolved independently according to the explicit form \eqref{eq:EOMComp}. Substitute Eq.~\eqref{eq:Invertibility} into Eq.~\eqref{eq:deciphereEqFull}, apply the product rule, and note that $S_{e}$ already satisfies Eq.~\eqref{eq:deciphereEqFull}, we must thus have $\mathcal{F}_{\mu}$ and $\mathcal{F}_{\tau}$, and subsequently $\mathbb{F}$, all being constants. 
\newline\newline
{\bf *8-} We motivate the decipherer to take on physically attractive forms in the particularly simple flat Minkowski spacetime. Such an endeavour will also provide us with boundary conditions for when integrating the decipherer into curved regions of an asymptotically flat spacetime. We begin by noting that  
each individual flavour component ${S}_{\hat{f}_0}$ is a transformation between spinor dyads, so in a flat spacetime\footnote{Also adopting Cartesian instead of curvilinear coordinates, so the associated dyad/tetrad has no dynamics itself (simply parallelly transported everywhere), and $a$ and $b$ are identity matrices. }, it is desirable to have them belong to $\text{SU}(2)$, thus map spin-frames in the particle spinor space into spin-frames in the spacetime spinor space, which physically means that $\mathring{S}_{\hat{f}_0}$ (overhead circles signify flat spacetime quantities) preserves orthonormality. 
We emphasis though that this is not an essential condition, only one based on aesthetics, because the spacetime dyad basis we actually use in computations (such as those in *5) is generally not the image of the particle dyad under $S$. 
We further note that an overall phase factor can be uniformly applied to all spacetime spinors, without changing the spacetime tensors that they map into (see Eq.~\ref{eq:SimpleMapping}, and we denote this freedom by $\text{U}_{\text{sp}}(1)$), so without compromising our aesthetics, the normalization 
\eqref{eq:Normalization} can be relaxed to pick up an overall phase factor on the spacetime spinor side, and thus each $\mathring{S}_{\hat{f}_0}$ (bridging a strict spin-frame on the particle side to a phase-relaxed spin-frame on the spacetime side) can be an $\text{SU}(2)$ matrix times an overall phase factor, or in other words, 
\bea \label{eq:FPh}
\mathcal{F}_{\mu} = e^{i\phi_\mu}\,, \quad 
\mathcal{F}_{\tau} = e^{i\phi_\tau}\,.
\eea
The base case $\mathring{S}_{e}$ can be written out explicitly as 
\bea \label{eq:SU2}
\mathring{S}_{e} = \bma 
\alpha & -\bar{\beta} \\
\beta & \bar{\alpha} 
\ema\,, \quad |\alpha|^2+|\beta|^2=1\,,
\eea
where the precise values of the complex constants $\alpha$ and $\beta$ are immaterial, as there is a representational spin-frame freedom in the particle spinor space (see *3), which allows us to overlay any arbitrary $\text{SU}(2)$ transformation on $\mathring{S}_{e}$ (and synchronously to other generations via Eq.~\eqref{eq:Invertibility}). We will assume $\alpha =\cos\zeta \in \mathbb{R}$ and $\beta =\sin\zeta \in \mathbb{R}$ for simplicity, and leave $\zeta$ free, so we could utilize the representational freedom to simplify calculations should such opportunities arise. 
With Eq.~\eqref{eq:FPh}, the projection operator $\mathbb{F}$ becomes weakly democratic in the sense that all its entries share the same absolute value. 

To pin down the phase angles $\phi_{\mu}$ and $\phi_{\tau}$, we note that since the physical mechanism for condensing three particle spinors of different flavours into a single spacetime spinor is through a projection in the flavour space, we would prefer $\gamma$ and more precisely its flavour constituent $\mathbb{G}$ to be a slightly generalized projection operator satisfying 
$\mathbb{G}^2=\pm \mathbb{G}$. With either sign, the normalization \eqref{eq:Normalization} of a spin-frame in the image of $\mathbb{G}$ will not be spoiled by subsequent projections. We do not allow for arbitrary phase increments during successive projections though, as $\mathbb{G}$ operates entirely within the particle side, and the $\text{U}_{\text{sp}}(1)$ freedom is not applicable. 
This requirement of $\mathbb{G}$ being a generalized projection operator is satisfied provided that
\bea \label{eq:ProjG}
e^{2i\phi_{\mu}}+e^{2i\phi_{\tau}} =\pm 1-1\,.
\eea
Condition \eqref{eq:ProjG} does not uniquely determine $\phi_{\mu}$ and $\phi_{\tau}$, so we keep these angles in their symbolic forms during derivations for generality. 
\newline\newline
{\bf *9-} We begin by noting that the metricity condition is an algebraic equation for $\Gamma$, and the first line constitutes the source terms. We can re-express them in the factorized form of 
\begin{align}
&\frac{\partial\gamma_{\alpha\, \hat{A}\, \hat{f}_0}{}^{\hat{A'}\, \hat{f}_1}}{\partial x^{\beta}}
- \Gamma^{\xi}_{\alpha \beta} \gamma_{\xi\, \hat{A}\, \hat{f}_0}{}^{\hat{A'}\, \hat{f}_1} \notag \\
=&\Big[\mathcal{P}_{\alpha }{}^{\eta}{}_{\beta}{}_{\hat{A}}{}^{\hat{E}}{}_{\hat{E'}}{}^{\hat{A'}}\Big]\Big[\mathbb{F}{}_{\hat{f}_0}{}^{\hat{f}_2}\bar{\mathbb{F}}{}_{\hat{f}_3}{}^{\hat{f}_1}\Big]\Big[\gamma_{\eta \, \hat{E}}{}^{\hat{E'}}{}_{\hat{f}_2}{}^{\hat{f}_3}\Big] \notag \\
=& \frac{1}{3}\Big[\mathcal{P}_{\alpha }{}^{\eta}{}_{\beta}{}_{\hat{A}}{}^{\hat{E}}{}_{\hat{E'}}{}^{\hat{A'}}\Big]\Big[\bar{\mathbb{F}}{}_{\hat{f}_3}{}^{\hat{f}_1}\mathbb{G}_{\hat{f}_2}{}^{\hat{f}_3}\mathbb{F}{}_{\hat{f}_0}{}^{\hat{f}_2}\Big]\Big[\gamma_{e\, \eta \, \hat{E}}{}^{\hat{E'}}\Big] \notag \\
=& \frac{1}{3}\Big[\mathcal{P}_{\alpha }{}^{\eta}{}_{\beta}{}_{\hat{A}}{}^{\hat{E}}{}_{\hat{E'}}{}^{\hat{A'}}\Big]\Big[\mathbb{G}_{\hat{f}_0}{}^{\hat{f}_1}\Big]\Big[\gamma_{e\, \eta \, \hat{E}}{}^{\hat{E'}}\Big]\,,
\end{align}
where using Eqs.~\eqref{eq:SComp}, \eqref{eq:SCompInv}, and \eqref{eq:Invertibility}, we obtain
\begin{align} \label{eq:DerivsTerms}
\mathcal{P}_{\alpha }{}^{\eta}{}_{\beta}{}_{\hat{A}}{}^{\hat{E}}{}_{\hat{E'}}{}^{\hat{A'}}=&\left(\frac{\partial b_{\alpha}{}^{\tilde{\delta}}}{\partial x^{\beta}}a^{\eta}{}_{\tilde{\delta}}-\Gamma^{\eta}_{\alpha\beta}\right)\epsilon_{\hat{A}}{}^{\hat{E}}\epsilon_{\hat{E'}}{}^{\hat{A'}} +\frac{\partial \bar{S}_e^{-1}{}_{\tilde{B'}}{}^{\hat{A'}}}{\partial x^{\beta}}\bar{S}_{e}{}^{\tilde{B'}}{}_{\hat{E'}}g_{\alpha}{}^{\eta}\epsilon_{\hat{A}}{}^{\hat{E}}-\frac{\partial S_e^{-1}{}_{\tilde{B}}{}^{\hat{E}}}{\partial x^{\beta}}
{S}_{e}{}^{\tilde{B}}{}_{\hat{A}}
g_{\alpha}{}^{\eta}\epsilon_{\hat{E'}}{}^{\hat{A'}}\,.
\end{align}
To make progress towards solving this equation, we note that if the flavour-reduced version of the metricity equation, namely 
\begin{align} \label{eq:MetricityFlavour2}
0=&\, \mathcal{P}_{\alpha }{}^{\eta}{}_{\beta}{}_{\hat{B}}{}^{\hat{E}}{}_{\hat{E'}}{}^{\hat{B'}}\gamma_{e\, \eta \, \hat{E}}{}^{\hat{E'}}+\bar{\Gamma}_{e \, \beta \, \hat{D'}}{}^{\hat{B'}}\gamma_{e\, \alpha \hat{B}}{}^{\hat{D'}}
-\gamma_{e\, \alpha \hat{C}}{}^{\hat{B'}} \Gamma_{e\, \beta\,  \hat{B}}{}^{\hat{C}} \,,
\end{align}
admits a solution, then 
\bea \label{eq:MetricityFlavour}
\Gamma_{\alpha \, \hat{B}\, \hat{f}_0}{}^{\hat{C}\, \hat{f}_1} = \Gamma_{e \, \alpha \, \hat{B}}{}^{\hat{C}}\mathbb{F}_{\hat{f}_0}{}^{\hat{f}_1}\,,  
\eea
would satisfy the full metricity condition by virtual of the identities \eqref{eq:FlavourIdentity}. 
Turning now to Eq.~\eqref{eq:MetricityFlavour2}, we further note that because $\Gamma$ enters into the equation linearly, we can make a decomposition
\bea
\Gamma_{e\, \beta\,  \hat{B}}{}^{\hat{C}} = {}^{\text{V}}\Gamma_{e\, \beta\,  \hat{B}}{}^{\hat{C}} +{}^{\text{S}}\Gamma_{e\, \beta\,  \hat{B}}{}^{\hat{C}} \,,
\eea
where 
\bea \label{eq:GammaNew}
{}^{\text{S}}\Gamma_{e\, \beta\,  \hat{A}}{}^{\hat{E}} = - \frac{\partial S_e^{-1}{}_{\tilde{B}}{}^{\hat{E}}}{\partial x^{\beta}}
{S}_{e}{}^{\tilde{B}}{}_{\hat{A}}\,,
\eea
balances the last two lines of Eq.~\eqref{eq:DerivsTerms}, leaving ${}^{\text{V}}\Gamma_{e}$ to take care of the first line, which is similar in form to the metricity source prior to the introduction of a dynamic decipherer. We can then lean on the standard technique in existing literature    to find ${}^{\text{V}}\Gamma_{e}$. 

One begins with the following second order elements of the Clifford algebra 
\begin{align}
\mathcal{B}_{\alpha\beta\, \hat{A}}{}^{\hat{C}} \equiv \frac{1}{2}\left\{\bar{\gamma}_{e\, \alpha \, \hat{D'}}{}^{\hat{C}}
\gamma_{e\, \beta \, \hat{A}}{}^{\hat{D'}}
-\left(\alpha \leftrightarrow \beta\right)\right\}\,,
\end{align}
to be used as basis to decompose ${}^{\text{V}}\Gamma_{e}$ into. Using Eq.~\eqref{eq:CliffordFull}, it is straight forward to verify that 
\begin{align}
&\bar{\mathcal{B}}_{\alpha\beta\, \hat{A'}}{}^{\hat{C'}}\gamma_{e\, \delta\, \hat{B}}{}^{\hat{A'}}-\gamma_{e\, \delta\, \hat{D}}{}^{\hat{C'}}{\mathcal{B}}_{\alpha\beta\, \hat{B}}{}^{\hat{D}}= g_{\alpha\delta}\gamma_{e\, \beta\, \hat{B}}{}^{\hat{C'}}-\left(\alpha \leftrightarrow \beta \right)\,.
\end{align}
Therefore, if we make the decomposition
\bea
{}^{\text{V}}\Gamma_{e\, \beta\, \hat{A}}{}^{\hat{C}} = \mathcal{Q}^{\delta \eta}{}_{\beta} \mathcal{B}_{\delta \eta\, \hat{A}}{}^{\hat{C}}\,,
\eea
and denote the term in the large bracket of the first line of Eq.~\eqref{eq:DerivsTerms} by ${}^{\text{V}}\mathcal{P}_{\alpha}{}^{\eta}{}_{\beta}$, then the $\text{V}$ segment of the metricity condition becomes an equation for $\mathcal{Q}$
\begin{align} 
\mathcal{Q}^{\delta \eta}{}_{\beta} \left(g_{\delta\alpha}\gamma_{\eta \, \hat{B}}{}^{\hat{B'}}-(\delta \leftrightarrow \eta)\right)= 2g_{\delta\alpha} \mathcal{Q}^{[\delta \eta]}{}_{\beta} \gamma_{\eta \, \hat{B}}{}^{\hat{B'}}=-{}^{\text{V}}\mathcal{P}_{\alpha }{}^{\delta}{}_{\beta}\gamma_{\delta \, \hat{B}}{}^{\hat{B'}}\,,
\end{align}
which is satisfied by 
\bea \label{eq:QandP}
\mathcal{Q}^{\eta\delta \beta} = -\frac{1}{2} {}^{\text{V}}\mathcal{P}^{[\eta\delta]\beta}\,.
\eea
\newline\newline
{\bf *10-} We write the metric in the Schwarzschild coordinates, which has the form
\bea \label{eq:SchwarzschildMetric}
ds^2= \frac{\Delta}{r^2} dt^2 -\frac{r^2}{\Delta}dr^2 - r^2 \left( d\theta^2 + \sin^2\theta d\phi^2 \right)\,, 
\eea
in geometrized units, where $\Delta = r(r-2M)$. 
The Newman-Penrose null tetrad we choose is the Kinnersley tetrad,
$\{\ell, n, m, \bar{m}\}$, 
\bea\label{eq:ExtKinn}
\ell^{\alpha} &=& \frac{1}{\Delta}\left(r^2,\Delta,0,0 \right)\,, \notag \\
n^{\alpha} &=& \frac{1}{2r^2}\left(r^2,-\Delta,0,0 \right)\,, \notag \\
m^{\alpha} &=& \frac{1}{\sqrt{2}r}\left(0,0,1,i \csc\theta \right)\,, 
\eea
which gives $b^{\alpha}{}_{\beta}$ when further combined with Eq.~\eqref{eq:Orthonormal}, with $b_{\beta}{}^{\alpha} = (T^{\alpha},N^{\alpha},E_2^{\alpha},E_3^{\alpha})$. Inverting $b_{\beta}{}^{\alpha}$, we get 
\bea \label{eq:aKinnersley}
a^{\alpha}{}_{\beta} = 
\left(
\begin{array}{cccc}
 \frac{3 r-2 M}{2 \sqrt{2} r} & -\frac{r(2 M+r)}{2 \sqrt{2} \Delta} & 0 & 0 \\
 -\frac{2 M+r}{2 \sqrt{2} r} & \frac{r(3 r-2 M)}{2 \sqrt{2} \Delta} & 0 & 0 \\
 0 & 0 & r & 0 \\
 0 & 0 & 0 & -r \sin \theta  \\
\end{array}
\right)\,.
\eea
The Ricci curvature for metric \eqref{eq:SchwarzschildMetric} vanishes and the nonvanishing entries in the Weyl curvature (modulo index symmetries) are 
\begin{align}
C_{0110} = -\frac{2M}{r^3}\,, \quad& 
C_{0220} =\frac{M\Delta}{r^3}\,, \notag \\
C_{2332} = 2Mr\sin^2\theta \,, \quad& 
C_{1221} = -\frac{Mr}{\Delta}\,, \notag \\
C_{1331} = C_{1221} \sin^2\theta\,, \quad& 
C_{0330} = C_{0220} \sin^2\theta\,.
\end{align}
The spin coefficients can be found in Ref.~\cite{ChandrasekharBook} Eqs.~21.287 and 21.288, with the nonvanishing ones being 
\bea
\rho = -\frac{1}{r}\,, &\quad& \beta= -\alpha=\frac{1}{2\sqrt{2}}\frac{\cot\theta}{r}\,, \notag \\
\mu = - \frac{\Delta}{2r^3}\,,&\quad& \gamma = \frac{M}{2r^2}\,.
\eea
On the other hand, the nonvanishing tensorial Christoffel symbols are (modulo symmetric partners)
\begin{align}\label{eq:ExtChris}
\Gamma^{0}_{10} &=-\Gamma^{1}_{11}= \frac{M}{\Delta}\,, \quad 
\Gamma^{1}_{00} = \frac{M\Delta}{r^4}\,, \quad 
\Gamma^{1}_{22} = -\frac{\Delta}{r}\,, \notag \\
\Gamma^{1}_{33} &= -\frac{\Delta\sin^2\theta}{r}\,, \quad 
\Gamma^{2}_{21} = \Gamma^{3}_{31} = \frac{1}{r}\,, \quad 
\Gamma^{2}_{33} = -\frac{1}{2}\sin2\theta\,, \notag \\
\Gamma^{3}_{32} &= \cot\theta\,. 
\end{align}
\newline\newline
{\bf *11-} We already know the flat spacetime solutions from (*8), albeit under a Cartesian tetrad adapted to a Cartesian coordinate system. We thus need to translate expressions there into (the $M \rightarrow 0$ limit of) the Kinnersley tetrad. Using overhead check to denote Cartesian quantities, we can write 
\bea
\mathring{S}_e^{-1}{}_{\tilde{D}}{}^{\hat{E}}=\mathring{\check{C}}^{\hat{E}}_{\o}\,\mathring{\check{\o}}_{\tilde{D}}+\mathring{\check{C}}^{\hat{E}}_{\iota}\,\mathring{\check{\iota}}_{\tilde{D}}\,.
\eea
Using Eq.~\eqref{eq:dyadExp} and the expression for $\mathring{S}_e$ from (*8), we obtain
\bea \label{eq:CartBC}
\mathring{\check{C}}_{\o}{}^{\hat{0}}=i\cos(\zeta -\pi/4)\,,&\,\,&
\mathring{\check{C}}_{\iota}{}^{\hat{0}}=\sin(\zeta -\pi/4)\,,\notag \\
\mathring{\check{C}}_{\o}{}^{\hat{1}}=i\sin(\zeta -\pi/4)\,,&\,\,&
\mathring{\check{C}}_{\iota}{}^{\hat{1}}=-\cos(\zeta -\pi/4)\,. 
\eea
We can also decompose $(\mathring{\check{\o}}_D,\mathring{\check{\iota}}_D)$ under the Kinnersley dyad $(\o_D,\iota_D)$ as 
\bea
\mathring{\check{\o}}_D=\mathring{P}_{\o}\,\mathring{\o}_D+\mathring{P}_{\iota}\,\mathring{\iota}_D\,,\quad
\mathring{\check{\iota}}_D=\mathring{Q}_{\o}\,\mathring{\o}_D+\mathring{Q}_{\iota}\,\mathring{\iota}_D\,.
\eea
To find these $\mathring{P}$ and $\mathring{Q}$ coefficients, we begin by finding the transformation $\mathcal{L}$ between the Cartesian and the Kinnersley null tetrads, 
\bea \label{eq:CartLorentz1}
\bma 
\mathring{\check{l}} \\
\mathring{\check{n}} \\
\mathring{\check{m}} \\
\mathring{\check{\bar{m}}} 
\ema = \mathcal{L}
\bma
\mathring{{l}} \\
\mathring{{n}} \\
\mathring{{m}} \\
\mathring{{\bar{m}}} 
\ema \,. 
\eea
Define the Cartesian coordinates 
\bea
x=r\sin\theta\cos\phi\,,\quad y=r\sin\theta\sin\phi\,, \quad z=r\cos\theta\,,  
\eea
and the corresponding orthonormal version of the Cartesian tetrad as $\{\mathring{\check{T}},\mathring{\check{N}},\mathring{\check{E}}_2,\mathring{\check{E}}_3\}=\{\partial_t, \partial_x,\partial_y, \partial_z\}$, then the Cartesian null tetrad obtained by inverting Eq.~\eqref{eq:Orthonormal} can be reached from the Kinnersley tetrad by an $\mathcal{L}$ equalling
\begin{align} \label{eq:CartLorentz2}
\left(
\begin{array}{cccc}
 \frac{c_{\phi }s_{\theta }+1}{2 \sqrt{2}} & \frac{1-c_{\phi } s_{\theta }}{\sqrt{2}} & \frac{ c_{\theta } c_{\phi }+i s_{\phi }}{2} & \frac{c_{\theta } c_{\phi }-i s_{\phi }}{2} \\
 \frac{-c_{\phi } s_{\theta }+1}{2 \sqrt{2}} & \frac{c_{\phi } s_{\theta }+1}{\sqrt{2}} & \frac{-c_{\theta } c_{\phi }-i s_{\phi }}{2} & \frac{i s_{\phi }-c_{\theta } c_{\phi }}{2} \\
 \frac{-c_{\theta }+i s_{\theta } s_{\phi }}{2 \sqrt{2}} & \frac{c_{\theta }-i s_{\theta } s_{\phi }}{\sqrt{2}} & \frac{c_{\phi }+s_{\theta }+i c_{\theta } s_{\phi }}{2} & \frac{-c_{\phi }+s_{\theta }+i c_{\theta } s_{\phi }}{2} \\
 \frac{-c_{\theta }-i s_{\theta } s_{\phi }}{2 \sqrt{2}} & \frac{c_{\theta }+i s_{\theta } s_{\phi }}{\sqrt{2}} & \frac{-c_{\phi }+s_{\theta }-i c_{\theta } s_{\phi }}{2} & \frac{c_{\phi }+s_{\theta }-i c_{\theta } s_{\phi }}{2} \\
\end{array}
\right)\,,
\end{align}
with $c_{\theta} \equiv \cos\theta$, $s_{\theta} \equiv \sin\theta$, $c_{\phi}\equiv \cos\phi$ and $s_{\phi}\equiv \sin\phi$. Its corresponding spin transformation is 
\bea
\mathcal{L}'=
\bma 
\mathring{P}_{\o} & \mathring{P}_{\iota} \\
\mathring{Q}_{\o} & \mathring{Q}_{\iota}\
\ema\,,
\eea
which has to be an element of $\text{SL}(2,\mathbb{C})$ that double covers the Lorentz group $\text{SO}(1,3)$, or in other words we must have
\bea \label{eq:DetSpinTrans}
\det \mathcal{L}'=\mathring{P}_{\o}\mathring{Q}_{\iota}-\mathring{P}_{\iota}\mathring{Q}_{\o}=1\,.
\eea
To explicitly find $\mathcal{L}'$, 
we can factorize the Cartesian null tetrad obtained from Eqs.~\eqref{eq:CartLorentz1} and \eqref{eq:CartLorentz2} into spinors using Eq.~\eqref{eq:DyadMap}, which leaves an overall phase freedom in $\mathcal{L}'$ that can be further pinned down with Eq.~\eqref{eq:DetSpinTrans}. The results are
\begin{align}\label{eq:Coeffs1}
\mathring{P}_{\o} =& \frac{\Lambda'}{4 \xi }
 \Bigg[4 \cos \theta  \cos \phi +i \left(\xi^2+\sqrt{2} (3 \sin \theta \cos \phi-1)\right)\Bigg]\,, \notag \\
\mathring{P}_{\iota} =& \frac{\Lambda'}{4 \xi} \Bigg[\xi^2+4 i \cos \theta \cos \phi +\sqrt{2} (1-3 \sin \theta \cos \phi)\Bigg]\,, \notag \\
\mathring{Q}_{\o} =& \frac{\Lambda' }{2 \xi } \Bigg[\sin \theta \left(\sqrt{2} \sin \phi +2\right)\notag \\
&-i \cos \theta \left(2 \sin \phi+\sqrt{2}\right)-2 \cos \phi \Bigg]\,, \notag \\
\mathring{Q}_{\iota} =& \frac{\Lambda'}{\xi} \Bigg[\cos \theta \left(\sin \phi +\sqrt{2}\right)\notag \\
&+i \left(\sqrt{2} \sin \theta \sin \phi +\sin \theta +\cos \phi \right)\Bigg]\,.
\end{align}
where 
\begin{align}\label{eq:Coeffs2}
\Lambda'=&\frac{\xi}{\sqrt{\chi}} \,, \quad
\xi = \sqrt{\sqrt{2} (3-\sin \theta \cos \phi)+4 \sin\phi}\,, \notag \\
\chi =& -4 \sin \theta \sin \phi +\sqrt{2} (\cos \phi -3 \sin \theta )+i \cos \theta  \left(3 \sqrt{2} \sin \phi +4\right)\,.
\end{align}
As a sanity check, we note that the Cartesian basis should be parallelly transported everywhere in a flat spacetime, so covariant derivatives such as $\delta \mathring{\check{\o}}_A$ should vanish, giving us 
\begin{align}
m^{\alpha}\partial_{\alpha} P_{\o}+\beta P_{\o}+\mu P_{\iota} &= 0\,, \notag \\ 
-m^{\alpha}\partial_{\alpha} P_{\iota}+\beta P_{\iota}+\sigma P_{\o} &= 0\,,
\end{align}
which is indeed satisfied by Eqs.~\eqref{eq:Coeffs1} and \eqref{eq:Coeffs2}. 
Substituting these and Eqs.~\eqref{eq:CartBC} into the Kinnersley dyad, we obtain
\begin{align}
\mathring{C}_{\o}{}^{\hat{E}} &= \mathring{\check{C}}_{\o}{}^{\hat{E}}\mathring{P}_{\o}+\mathring{\check{C}}_{\iota}{}^{\hat{E}}\mathring{Q}_{\o}\,, \notag \\
\mathring{C}_{\iota}{}^{\hat{E}} &= \mathring{\check{C}}_{\o}{}^{\hat{E}}\mathring{P}_{\iota}+\mathring{\check{C}}_{\iota}{}^{\hat{E}}\mathring{Q}_{\iota}\,, 
\end{align}
which we explicitly verify to satisfy the $M\rightarrow 0$ limit of the DEOM. 
The full curved-spacetime DEOM then reduces to
\begin{align} \label{eq:EOMExtPert}
\frac{\cot \theta \delta^{\hat{E}}_{\iota}}{2 \sqrt{2}}+\frac{1}{\sqrt{2}}\frac{\partial \delta^{\hat{E}}_{\iota}}{\partial \theta}-\frac{i\csc\theta}{\sqrt{2}}\frac{\partial \delta^{\hat{E}}_{\iota}}{\partial \phi}+\delta^{\hat{E}}_{\o}+r \frac{\partial\delta^{\hat{E}}_{\o} }{\partial r} &= 0\,, \notag \\
4 M \frac{\partial \mathring{C}^{\hat{E}}_{\iota}}{\partial r}+\frac{2 M}{r} \mathring{C}^{\hat{E}}_{\iota}
-2 \frac{r-M}{r} \delta^{\hat{E}}_{\iota}-2\frac{\Delta}{r} \frac{\partial \delta^{\hat{E}}_{\iota} }{\partial r}+\sqrt{2} \cot \theta  \delta^{\hat{E}}_{\o} +2\sqrt{2}i \csc\theta \frac{\partial \delta^{\hat{E}}_{\o}}{\partial \phi}+2\sqrt{2}\frac{\partial \delta^{\hat{E}}_{\o}}{\partial \theta}
&=0 \,,
\end{align}
that govern $\delta^{\hat{E}}_{\o/\iota}$, with $\mathring{C}^{\hat{E}}_{\o/\iota}$ serving as source terms.  
\newline\newline
{\bf *12-} To evaluate the integral in the solution for ${f}^{\hat{E}}_{\o}$, we begin by noting that the expression for $\mathring{C}^{\hat{E}}_{\iota}$ is simplified if we select $\zeta=\pi/4$ (see *8). However, it is still too complicated to be integrated analytically, so we opt for the numerical alternative. Taking $\theta = \pi/2$ and $\phi=0$ as an example angular location, we numerically evaluate the integrals in ${f}^{\hat{E}}_{\o}(\pi/2,0)$ (using $\mathcal{I}^{\hat{E}}(\theta',\theta,\phi)$ to denote the integrand), as well as  

\begin{figure}[t!]
  \centering
\begin{overpic}[width=0.24\columnwidth]  {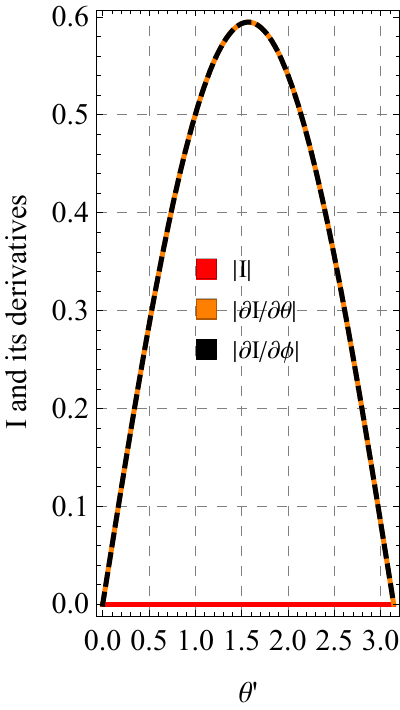}
\put(45,0){(a)}
\end{overpic}
\begin{overpic}[width=0.24\columnwidth]  {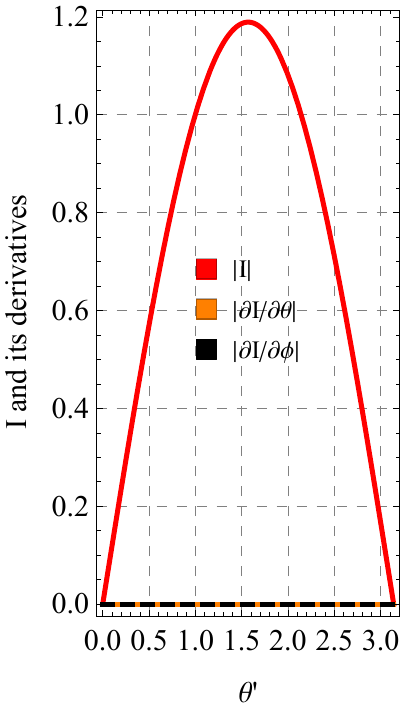}
\put(45,0){(b)}
\end{overpic}
  \caption{The absolute values $|\mathcal{I}^{\hat{E}}|$, $|\partial_{\theta}\mathcal{I}^{\hat{E}}|$ and $|\partial_{\phi}\mathcal{I}^{\hat{E}}|$ as functions of $\theta'$ evaluated at $\theta=\pi/2$ and $\phi=0$. 
  (a): for particle spinor index $\hat{E}=\hat{0}$. (b): for $\hat{E}=\hat{1}$.
}
	\label{fig:ExtIntegrand}
\end{figure}

\begin{align}
\frac{\partial {f}^{\hat{E}}_{\o}}{\partial \theta}\left(\frac{\pi}{2},0\right) =& -\frac{M}{\sqrt{2}} \Big[\int_1^{\frac{\pi }{2}} \frac{\partial \mathcal{I}^{\hat{E}}\left(\theta',\theta,0\right)}{\partial \theta}\Big|_{\theta=\pi/2} \, d\theta' +\mathcal{I}^{\hat{E}}\left(\frac{\pi }{2},\frac{\pi }{2},0\right)\Big]
+\frac{\partial\mathfrak{g}^{\hat{E}}_{\o}}{\partial \theta}\Big|_{\theta=\pi/2,\, \phi=0}
\,, \notag \\
\frac{\partial {f}^{\hat{E}}_{\o}}{\partial \phi}\left(\frac{\pi}{2},0\right) =& -\frac{M}{\sqrt{2}} \int_1^{\frac{\pi }{2}} \frac{\partial \mathcal{I}^{\hat{E}}\left(\theta',\pi/2,\phi\right)}{\partial \phi}\Big|_{\phi=0} \, d\theta' +\frac{\partial\mathfrak{g}^{\hat{E}}_{\o}}{\partial \phi}\Big|_{\theta=\pi/2,\,\phi=0}
\,,
\end{align}
that are needed in later calculations. We plot the $\mathcal{I}^{\hat{E}}$, $\partial_{\theta}\mathcal{I}^{\hat{E}}$ and $\partial_{\phi}\mathcal{I}^{\hat{E}}$ values at $\theta=\pi/2$ and $\phi=0$ in Fig.~\ref{fig:ExtIntegrand}, which after integration gives 
\begin{align}
f^{\hat{0}}_{\o}\left(\frac{\pi}{2},0\right)  &= \mathfrak{g}^{\hat{0}}_{\o}\Big|_{\theta=\pi/2,\, \phi=0} \,, \notag \\
\frac{\partial f^{\hat{0}}_{\o}}{\partial \theta}\left(\frac{\pi}{2},0\right) &= \frac{\partial\mathfrak{g}^{\hat{0}}_{\o}}{\partial \theta}\Big|_{\theta=\pi/2,\, \phi=0}+\lambda_{\text{ex}} \frac{i M}{\sqrt{2}} 
\,, \notag \\
\frac{\partial f^{\hat{0}}_{\o}}{\partial \phi}\left(\frac{\pi}{2},0\right) &= \frac{\partial\mathfrak{g}^{\hat{0}}_{\o}}{\partial \phi}\Big|_{\theta=\pi/2,\, \phi=0}-\lambda_{\text{ex}}\frac{M}{\sqrt{2}}\,,
\end{align}
where $\lambda_{\text{ex}}=0.3213$, and 
\begin{align}
f^{\hat{1}}_{\o}\left(\frac{\pi}{2},0\right)  &= \mathfrak{g}^{\hat{1}}_{\o}\Big|_{\theta=\pi/2,\, \phi=0} + 2\lambda_{\text{ex}}\frac{M}{\sqrt{2}}
\,, \notag \\
\frac{\partial f^{\hat{1}}_{\o}}{\partial \theta}\left(\frac{\pi}{2},0\right) &= \frac{\partial\mathfrak{g}^{\hat{1}}_{\o}}{\partial \theta}\Big|_{\theta=\pi/2,\, \phi=0}
\,, \notag \\
\frac{\partial f^{\hat{1}}_{\o}}{\partial \phi}\left(\frac{\pi}{2},0\right) &= \frac{\partial\mathfrak{g}^{\hat{1}}_{\o}}{\partial \phi}\Big|_{\theta=\pi/2,\, \phi=0}\,.
\end{align}
\newline\newline
{\bf *13-} Define
\bea
\acute{\Gamma}^{\eta}_{\alpha \beta} \equiv
a^{\eta}{}_{\tilde{\delta}}\frac{\partial b_{\alpha}{}^{\tilde{\delta}}}{\partial x^{\beta}}\,,
\eea
then its non-vanishing components are (using Eq.~\ref{eq:aKinnersley})
\begin{align}
\acute{\Gamma}^{0}_{01} &= -\acute{\Gamma}^{1}_{11}  = -M\left(\frac{1}{4r^2}+\frac{r^2}{\Delta^2}\right)\,, \notag \\ 
\acute{\Gamma}^{0}_{11} &= \frac{M(4M^2+4Mr-7r^2)}{4\Delta^2}\,, \notag \\
\acute{\Gamma}^{1}_{01} &= -\frac{M(4M^2-12Mr+r^2)}{4\Delta^2}\,, \notag \\
\acute{\Gamma}^{2}_{21} &=\acute{\Gamma}^{3}_{31} =\frac{1}{r}\,, \quad 
\acute{\Gamma}^{3}_{32} = -\cot\theta\,.
\end{align}
Together with Eq.~\eqref{eq:ExtChris}, these values give us ${}^{\text{V}}\Gamma_{e\, \alpha}{}_{\hat{B}}{}^{\hat{A}}$. 

Piecing together the fragments from the main text and (*12), we obtain the following explicit expressions for the decipherer and its nonvanishing derivatives at the location $\theta=\pi/2$, $\phi=0$ (and arbitrary $r$):
\begin{align} \label{eq:SDerivs}
S_e^{-1}{}_{\tilde{D}}{}^{\hat{E}}&=
\frac{1}{\sqrt{2} r}
\left(
\begin{array}{cc}
- \mathfrak{g}^{\hat{0}}_{\iota}- i \mathfrak{g}^{\hat{0}}_{\o}+2^{-1/4} r & 
- \mathfrak{g}^{\hat{0}}_{\iota}- i \mathfrak{g}^{\hat{0}}_{\o}+2^{-1/4} r \\
 2^{1/4} r-i \left(\mathfrak{g}^{\hat{1}}_{\o}+\mathfrak{g}^{\hat{1}}_{\iota} (-i)+\sqrt{2} M \lambda_{\text{ex}}\right) & 2^{1/4} r+i \left(\mathfrak{g}^{\hat{1}}_{\o}+\mathfrak{g}^{\hat{1}}_{\iota} i+\sqrt{2} M \lambda_{\text{ex}}\right) \\
\end{array}
\right)
\,,\notag \\
S_e{}^{\tilde{D}}{}_{\hat{E}}&=
\frac{1}{N_1}
\left(
\begin{array}{cc}
 \sqrt{2} r \left(\mathfrak{g}^{\hat{1}}_{\iota}-i \left(\mathfrak{g}^{\hat{1}}_{\o}-i 2^{1/4} r+\sqrt{2} M \lambda_{\text{ex}}\right)\right) & 
 -r \left(2 \mathfrak{g}^{\hat{0}}_{\iota}-2 i \mathfrak{g}^{\hat{0}}_{\o}+2^{3/4} r\right) \\
 \sqrt{2} r \left(-\mathfrak{g}^{\hat{1}}_{\iota}-i \mathfrak{g}^{\hat{1}}_{\o}+2^{1/4} r-i \sqrt{2} M \lambda_{\text{ex}}\right) & 
 -r \left(-2 \mathfrak{g}^{\hat{0}}_{\iota}-2 i \mathfrak{g}^{\hat{0}}_{\o}+2^{3/4} r\right) \\
\end{array}
\right)
\,,\notag \\
N_1 &\equiv {\mathfrak{g}^{\hat{1}}_{\iota} \left(-2 i \mathfrak{g}^{\hat{0}}_{\o}+2^{3/4} r\right)+2 i \left(r \left(2^{1/4} \mathfrak{g}^{\hat{0}}_{\o}+i r\right)+\mathfrak{g}^{\hat{0}}_{\iota} \left(\mathfrak{g}^{\hat{1}}_{\o}+\sqrt{2} M \lambda_{\text{ex}}\right)\right)}\,, \notag \\
\frac{\partial S_e^{-1}{}_{\tilde{D}}{}^{\hat{E}}}{\partial r}&=
\frac{1}{\sqrt{2} r^2}\left(
\begin{array}{cc}
 \mathfrak{g}^{\hat{0}}_{\iota}+\mathfrak{g}^{\hat{0}}_{\o} i & \mathfrak{g}^{\hat{0}}_{\iota}-i \mathfrak{g}^{\hat{0}}_{\o} \\
 (\mathfrak{g}^{\hat{1}}_{\iota}+\mathfrak{g}^{\hat{1}}_{\o} i)+\sqrt{2} i M \lambda_{\text{ex}} &  (\mathfrak{g}^{\hat{1}}_{\iota}-i \mathfrak{g}^{\hat{1}}_{\o})-\sqrt{2} i M \lambda_{\text{ex}} \\
\end{array}
\right)
\,,\notag \\
\frac{\partial S_e^{-1}{}_{\tilde{D}}{}^{\hat{E}}}{\partial \theta}&=
\frac{1}{4r}\left(
\begin{array}{cc}
 -2 \sqrt{2} (\mathfrak{g}^{\hat{0}'}_{\o}+\mathfrak{g}^{\hat{0}'}_{\iota} i)+2^{3/4} i r+2 M \lambda_{\text{ex}} & i 2^{3/4} r-2 \left(i \sqrt{2} (\mathfrak{g}^{\hat{0}'}_{\iota}+\mathfrak{g}^{\hat{0}'}_{\o} i)+M \lambda_{\text{ex}}\right) \\
i \left(-2\sqrt{2} \mathfrak{g}^{\hat{1}'}_{\iota}+2\sqrt{2} \mathfrak{g}^{\hat{1}'}_{\o} i+2^{1/4} r\right) & -i \left(2\sqrt{2} \mathfrak{g}^{\hat{1}'}_{\iota}+2\sqrt{2} \mathfrak{g}^{\hat{1}'}_{\o} i+2^{1/4} r\right) \\
\end{array}
\right)
\,,\notag \\
\frac{\partial S_e^{-1}{}_{\tilde{D}}{}^{\hat{E}}}{\partial \phi}&=
-\frac{1}{2r}
\left(
\begin{array}{cc}
\sqrt{2} (\mathfrak{g}^{\hat{0}'}_{\iota}+\mathfrak{g}^{\hat{0}'}_{\o} i)+2^{-1/4} r-i M \lambda_{\text{ex}} &  \sqrt{2} (\mathfrak{g}^{\hat{0}'}_{\iota}-i \mathfrak{g}^{\hat{0}'}_{\o})+2^{-1/4} r+ i M \lambda_{\text{ex}} \\
\sqrt{2} \mathfrak{g}^{\hat{1}'}_{\iota}+\sqrt{2} i \mathfrak{g}^{\hat{1}'}_{\o}-2^{-3/4} r & \sqrt{2} \mathfrak{g}^{\hat{1}'}_{\iota}-\sqrt{2} i \mathfrak{g}^{\hat{1}'}_{\o}+2^{-3/4} r \\
\end{array}
\right)
\,.\notag \\
\end{align}
Defining
\bea
\mathfrak{p}_1\equiv \mathfrak{g}^{\hat{1}}_{\o}+\sqrt{2} M \lambda_{ex}\,, \quad \mathfrak{p}_2\equiv M \lambda_{ex}- \sqrt{2} \mathfrak{g}^{\hat{0'}}_{\o}\,, \quad
\mathfrak{p}_3\equiv 2^{1/4} r-\mathfrak{g}^{\hat{1}}_{\iota}\,, \quad 
\mathfrak{p}_4\equiv  r-2^{1/4}i\mathfrak{g}^{\hat{0}}_{\o}\,,
\eea
the spin connection components ${}^{\text{S}}\Gamma_{e\, \alpha}{}_{\hat{B}}{}^{\hat{A}}$ are then simply 
\begin{align}
-\frac{\partial S_e^{-1}{}_{\tilde{B}}{}^{\hat{E}}}{\partial r}
{S}_{e}{}^{\tilde{B}}{}_{\hat{A}}&=\frac{1}{N_1}
\left(
\begin{array}{cc}
 \frac{2 i}{r}  \left(\mathfrak{p}_3 \mathfrak{g}^{\hat{0}}_{\o}+\mathfrak{g}^{\hat{0}}_{\iota} \mathfrak{p}_1\right)& 2^{3/4} \mathfrak{g}^{\hat{0}}_{\iota} \\
 i 2^{5/4} \mathfrak{p}_1 & 
\frac{2^{3/4}}{r}\mathfrak{p}_4 \mathfrak{g}^{\hat{1}}_{\iota}+\frac{2 i}{r}\mathfrak{g}^{\hat{0}}_{\iota} \mathfrak{p}_1 \\
\end{array}
\right)
\,,\notag \\
-\frac{\partial S_e^{-1}{}_{\tilde{B}}{}^{\hat{E}}}{\partial \theta}
{S}_{e}{}^{\tilde{B}}{}_{\hat{A}}&=\frac{-1}{N_1}
\left(
\begin{array}{cc}
 2^{1/4} (2 \mathfrak{g}^{\hat{0}'}_{\o}+\mathfrak{g}^{\hat{1}}_{\o}) r+\sqrt{2}\mathfrak{g}^{\hat{1}}_{\iota} \mathfrak{p}_2-2 \mathfrak{g}^{\hat{0}'}_{\iota} \mathfrak{p}_1 & -\sqrt{2} \mathfrak{g}^{\hat{0}}_{\iota} \mathfrak{p}_2-i2^{-1/4} \mathfrak{p}_4 \left(2^{1/4} r-2 \mathfrak{g}^{\hat{0}'}_{\iota}\right) \\
 -2^{-1/4} i r \mathfrak{p}_3+2 \mathfrak{g}^{\hat{1}'}_{\o} \mathfrak{p}_3-2 \mathfrak{g}^{\hat{1}'}_{\iota} \mathfrak{p}_1 & 2^{-1/4} \left(2 i\mathfrak{g}^{\hat{1}'}_{\iota} \mathfrak{p}_4+2^{5/4} \mathfrak{g}^{\hat{0}}_{\iota} \mathfrak{g}^{\hat{1}'}_{\o}-i \mathfrak{g}^{\hat{0}}_{\iota} r \right) \\
\end{array}
\right)\,,\notag \\
-\frac{\partial S_e^{-1}{}_{\tilde{B}}{}^{\hat{E}}}{\partial \phi}
{S}_{e}{}^{\tilde{B}}{}_{\hat{A}}&=\frac{-1}{N_1}
\left(
\begin{array}{cc}
 i \left(2^{1/4} (2 \mathfrak{g}^{\hat{0}'}_{\o}+\mathfrak{g}^{\hat{1}}_{\o}) r+\sqrt{2}\mathfrak{g}^{\hat{1}}_{\iota} \mathfrak{p}_2+2 \mathfrak{g}^{\hat{0}'}_{\iota} \mathfrak{p}_1\right) 
 & -\sqrt{2}i \mathfrak{g}^{\hat{0}}_{\iota} \mathfrak{p}_2+2^{-1/4} \mathfrak{p}_4 \left(2 \mathfrak{g}^{\hat{0}'}_{\iota}+2^{1/4} r\right) \\
-2^{-1/4} r \mathfrak{p}_3+2 i \mathfrak{g}^{\hat{1}'}_{\o} \mathfrak{p}_3+2 i\mathfrak{g}^{\hat{1}'}_{\iota}  \mathfrak{p}_1 & 2^{-1/4} \left(2 \mathfrak{g}^{\hat{1}'}_{\iota} \mathfrak{p}_4 + 2^{5/4} i\mathfrak{g}^{\hat{0}}_{\iota} \mathfrak{g}^{\hat{1}'}_{\o}-\mathfrak{g}^{\hat{0}}_{\iota} r\right) \\
\end{array}
\right)\,.
\end{align}
\newline\newline
{\bf *14-} The neutrino oscillation phenomenology is conventionally described with an $3\times 3$ unitary matrix (if one does not wish to commit to the intrinsic mass explanation for neutrino flavour oscillations, they should regard the quantities below as bookkeeping tools for organizing experimental data) called the Pontecorvo-Maki-Nakagawa-Sakata (PMNS) matrix $U_{i}{}^{\hat{f}_0}$ (where lower case Latin letters from the middle of the alphabet label ``mass eigenstates''), which is usually written in the Particle Data Group convention as (leaving out the two Majorana CP violation angles that have no implications for oscillations) 
\begin{align} \label{eq:PMNSFull}
\bma 
1 & 0 & 0\\
0 & c_{23} & s_{23} \\
0 & -s_{23} & c_{23} 
\ema
\bma 
c_{13} & 0 & s_{13}e^{-i\delta}\\
0 & 1 & 0 \\
-s_{13}e^{i\delta} & 0 & c_{13} 
\ema
\bma 
c_{12} & s_{12} & 0\\
-s_{12} & c_{12} & 0\\
0 & 0 & 1
\ema
\,.
\end{align}
The symbols $c_{ij}\equiv \cos(\theta_{ij})$ and $s_{ij}\equiv \sin(\theta_{ij})$ denote trigonometrical functions of the mixing angles $0\leq \theta_{ij}\leq \pi/2$. Another piece of the jigsaw are the masses $m_i$ of the mass eigenstates. If we define ($L$ being the oscillation baseline distance and $E$ the neutrino energy)
\bea
M_{i}{}^{j} = 
\bma e^{-i m^2_1 L/2E} & 0 & 0 \\
0 & e^{-i m^2_2 L /2E} & 0 \\
0 & 0 & e^{-i m^2_3 L /2E} 
\ema\,, 
\eea
then the transition probability from $\hat{f}_0$ to $\hat{f}_1$ for ultra-relativistic neutrinos is given by 
\bea
P_{\hat{f}_0 \rightarrow \hat{f}_1} =|U_{j}{}^{\hat{f}_1}M_{i}{}^{j}U^{\dagger}_{\hat{f}_0}{}^{i}|^2\,,
\eea
which for an illustrative two-flavour case becomes 
\begin{align} \label{eq:IMDTransProb}
&P_{\nu_e\rightarrow \nu_{\mu}}=\sin^2(2\theta_{12})\sin^2 \frac{\Delta m_{21}^2 L}{4E}\,, \notag \\ 
&\Delta m_{21}^2 \equiv m^2_2-m^2_1\,.
\end{align}
The current best estimates for the parameters appearing above, as inferred from the experimentally measured transition probabilities, are \cite{Olive:2016xmw} (values for the inverted mass hierarchy are shown in the square brackets)
\begin{align} \label{eq:measured}
&\sin^2 (\theta_{12}) \approx 0.304 \pm 0.014\,, \notag \\
&\sin^2 (\theta_{13}) \approx (2.19\pm 0.12) \times 10^{-2}\,,\notag \\
&\sin^2 (2\theta_{23}) \approx 0.51 [0.50] \pm 0.05\,, \notag \\
&\Delta m^2_{21} \approx (7.53\pm 0.18)\times 10^{-5} \text{eV}^2 \,, \notag \\
&\Delta m^2_{31} \approx \Delta m^2_{32} \approx (2.44[2.51]\pm 0.06)\times 10^{-3} \text{eV}^2 \,,
\end{align}
and present indications are consistent with $\delta \approx 3\pi/2$ \cite{Abe:2017uxa}. 

The underlying flavour evolution leading to the aforementioned transition probabilities is of the phenomenological form 
\bea \label{eq:HamiltonianEvolution}
i\frac{d}{d\tau}\bma \nu_e \\ \nu_{\mu} \\ \nu_{\tau} \ema = \mathbb{H}(\tau)\bma \nu_e \\ \nu_{\mu} \\ \nu_{\tau} \ema \,,
\eea
where the $\nu$s represent flavour eigenstates, and the Hamiltonian $\mathbb{H}$ is given by 
\bea
\mathbb{H}_{\hat{f}_0}{}^{\hat{f}_1} = U_{j}{}^{\hat{f}_1}\mathcal{M}_{i}{}^{j} U^{\dagger}_{\hat{f}_0}{}^{i}\,,
\eea
where
\bea \label{eq:MassMatrix}
\mathcal{M}_{i}{}^{j} = \bma
\frac{m^2_1}{2E} & 0 & 0 \\
0 & \frac{m^2_2}{2E} & 0 \\
0 & 0 & \frac{m^2_3}{2E}
\ema\,.
\eea
In other words, an integral version of Eq.~\eqref{eq:HamiltonianEvolution} would simply consists of converting flavour eigenstates to mass eigenstates, evolving their phases according to their respective masses, and then converting back to flavours. An equal shift of all the phases will not affect flavour oscillations, so we can take $m^2_1/(2E)\mathbb{1}_i{}^j$ off Eq.~\eqref{eq:MassMatrix} to obtain an equivalent
\bea
\mathcal{M}_{i}{}^{j} = \bma
0 & 0 & 0 \\
0 & \frac{\Delta m^2_{21}}{2E} & 0 \\
0 & 0 & \frac{\Delta m^2_{31}}{2E}
\ema\,.
\eea
Starting from the following set of slightly idealized experimental values (assuming there exists complications, as yet unaccounted for, that nudge the measured mixing parameters off the values of the  more fundamental underlying quantities): $\Delta m_{21} \sim 0$ (since $\Delta m_{21} \ll \Delta m_{31} $, with the consequence that $\theta_{12}$ does not contribute to $\mathbb{H}$ below), $\theta_{23}=\pi/4$ (ensures that the lower right corner is a democratic $2\times 2$ matrix), and $\delta=3\pi/2$ (fixing the phase angle of the complex entries), we get 
\bea
\mathbb{H}_{\hat{f}_0}{}^{\hat{f}_1} \propto \frac{1}{3}\bma 2 t^2_{13} & i \sqrt{2} t_{13} &  i \sqrt{2} t_{13} \\
-i \sqrt{2} t_{13} & 1 & 1 \\
-i \sqrt{2} t_{13} & 1 & 1
\ema\,, 
\eea
where $t_{13}\equiv \tan\theta_{13}$. 
If we further set $\theta_{13}$ to be equal to the idealized $\theta_{12}=\arcsin(1/\sqrt{3})$, then $\mathbb{H}$ becomes a projection operator having the same form \eqref{eq:FDef} as $\mathbb{F}$, with $\phi_{\mu}=\phi_{\tau}=\pi/2$.

\end{document}